\documentclass[journal]{IEEEtran}

\usepackage{cite}
\usepackage{amsmath,amssymb,amsfonts,amsthm}
\usepackage{algorithmic}
\usepackage{graphicx}
\usepackage{textcomp}
\usepackage{xcolor}
\usepackage{hyperref}
\usepackage[ruled]{algorithm2e}
\usepackage{autobreak}
\usepackage{multirow}
\hypersetup{hidelinks}
\newtheorem{remark}{Remark}
\def\Ab{\mathbf{A}}
\def\ab{\mathbf{a}}

\def\bb{\mathbf{b}}

\def\Db{\mathbf{D}}

\def\Hb{\mathbf{H}}
\def\hb{\mathbf{h}}
\def\Ib{\mathbf{I}}

\def\mb{\mathbf{m}}

\def\nb{\mathbf{n}}

\def\Qb{\mathbf{Q}}
\def\qb{\mathbf{q}}

\def\rb{\mathbf{r}}

\def\ub{\mathbf{u}}
\def\Vb{\mathbf{V}}
\def\vb{\mathbf{v}}
\def\Wb{\mathbf{W}}
\def\wb{\mathbf{w}}
\def\Xb{\mathbf{X}}
\def\xb{\mathbf{x}}

\def\yb{\mathbf{y}}

\def\zb{\mathbf{z}}

\def\Cbb{\mathbb{C}}
\def\Rbb{\mathbb{R}}
\def\Ebb{\mathbb{E}}
\def\Zbb{\mathbb{Z}}

\def\x{\times}

\begin{document}

\title{Scalable Transceiver Design for Multi-User Communication in FDD Massive MIMO Systems via Deep Learning}

\author{Lin~Zhu,
        Weifeng~Zhu,
        Shuowen~Zhang,
        Shuguang~Cui,
        and~Liang~Liu% <-this % stops a space
\thanks{L. Zhu, W. Zhu, S. Zhang, and L. Liu are with the Department of Electrical and Electronic Engineering, The Hong Kong Polytechnic University, Hong Kong SAR (email: \href{lin-eee.zhu@connect.polyu.hk}{lin-eee.zhu@connect.polyu.hk}, \{\href{eee-wf.zhu@polyu.edu.hk}{eee-wf.zhu}, \href{shuowen.zhang@polyu.edu.hk}{shuowen.zhang}, \href{liang-eie.liu@polyu.edu.hk}{liang-eie.liu}\}@polyu.edu.hk).}
\thanks{S. Cui is with the School of Science and Engineering and the Future Network of Intelligence Institute, The Chinese University of Hong Kong, Shenzhen, 518172, China (email: \href{shuguangcui@cuhk.edu.cn}{shuguangcui@cuhk.edu.cn}).}
}% <-this % stops a space}

\maketitle

\begin{abstract}
This paper addresses the joint transceiver design, including pilot transmission, channel feature extraction and feedback, as well as precoding, for low-overhead downlink massive multiple-input multiple-output (MIMO) communication in frequency-division duplex (FDD) systems. Although deep learning (DL) has shown great potential in tackling this problem, existing methods often suffer from poor scalability in practical systems, as the solution obtained in the training phase merely works for a fixed feedback capacity and a fixed number of users in the deployment phase. To address this limitation, we propose a novel DL-based framework comprised of choreographed neural networks, which can utilize one training phase to generate all the transceiver solutions used in the deployment phase with varying sizes of feedback codebooks and numbers of users. The proposed framework includes a residual vector-quantized variational autoencoder (RVQ-VAE) for efficient channel feedback and an edge graph attention network (EGAT) for robust multi-user precoding. It can adapt to different feedback capacities by flexibly adjusting the RVQ codebook sizes using the hierarchical codebook structure, and scale with the number of users through a feedback module sharing scheme and the inherent scalability of EGAT. Moreover, a progressive training strategy is proposed to further enhance data transmission performance and generalization capability. Numerical results on a real-world dataset demonstrate the superior scalability and performance of our approach over existing methods.
\end{abstract}


\IEEEpeerreviewmaketitle

\section{Introduction}
%\IEEEPARstart{F}{or} 
Massive multiple-input multiple-output (MIMO) has been recognized as an indispensable technology for future wireless communication systems, offering substantial improvements in both capacity and reliability performance \cite{Larsson_MCOM, Lulu_JSTSP}. By leveraging the significantly large spatial degrees of freedom provided by massive MIMO, it is possible to serve a greater number of users simultaneously, thereby maximizing the benefits of multiplexing gains. To fully realize these advantages, low-overhead channel state information (CSI) acquisition is crucial, especially for downlink communication. In time-division duplex (TDD) systems, by exploiting the reciprocity between uplink and downlink channels, downlink CSI can be efficiently estimated through uplink pilots, whose overhead depends on the large number of antennas at the base station (BS). However, in frequency-division duplex (FDD) systems, how to acquire downlink CSI with low overhead for subsequent BS precoding design remains an open problem. 

This paper considers the downlink multi-user communications in FDD massive MIMO systems. Due to the huge number of BS antennas, the conventional schemes designed for systems with a small number of BS antennas \cite{Jindal-TIT, Jindal-JSAC, XiaPengfei_TSP}, under which the three core modules---channel estimation, CSI feedback, and multi-user precoding---are designed independently, will lead to unacceptable overhead. To address this limitation, the joint design of CSI estimation and feedback as well as multi-user precoding is considered for the transceiver design. Since the joint design introduces a complicated functional optimization problem that is usually infeasible to solve under traditional optimization techniques, the deep learning (DL) techniques have been widely used to solve this problem \cite{Yu-TWC, GaoZhen_JSAC, HoonLee-TCom, Caire-ICC, Carpi-ICC}. However, existing DL-based approaches suffer from poor generalization and typically require extensive retraining when system parameters such as feedback capacities or user numbers vary, which hinders their practical implementation in dynamic environments. Consequently, the scalable DL-based transceiver framework design remains an open problem. Once trained on a specific dataset, such a framework should deliver consistently high performance across varying system configurations.

\subsection{Prior Works}
Driven by the success of DL techniques in many areas, numerous studies have proposed neural network (NN)-based solutions for reducing the overhead of channel estimation and channel feedback in downlink massive MIMO systems \cite{HuangHongji, DongPeihao, CSINet}. Specifically, \cite{HuangHongji} introduces a DL-based framework for channel estimation and direction-of-arrival estimation in massive MIMO, utilizing NNs to capture the statistical properties of wireless channels and their spatial structure in the angular domain. Similarly, \cite{DongPeihao} proposes a deep convolutional neural network (CNN)-based channel estimator that leverages the temporal correlations inherent in time-varying channels. In the context of CSI feedback, \cite{CSINet} adopts an autoencoder architecture, where the encoder is deployed at the user to compress the channel matrix into a low-dimensional vector, and the decoder is placed at the BS to reconstruct the channel from the compressed vector. Although these works can reduce CSI acquisition overhead to some extent, their objective of accurately reconstructing full CSI at the BS inevitably leads to overhead that is still proportional to the huge number of BS antennas. 

Alternatively, the works \cite{Yu-TWC, GaoZhen_JSAC} point out that an implicit feature vector capturing essential characteristics of each user's high-dimensional channel vector may be sufficient for the BS to design its precoding matrix for downlink communication. Recognizing that the optimal channel feature and the BS precoding design are highly dependent, different from \cite{HuangHongji, DongPeihao, CSINet, XiaWenchao} that focus on independent module design, these works propose to utilize the DL technique to jointly design the BS pilot signals, user CSI extraction and feedback, and BS precoding. Specifically, the DL frameworks in \cite{Yu-TWC, GaoZhen_JSAC} utilize NNs to optimize pilot signals that can better explore the wireless environment. For CSI extraction and feedback, a fully connected network (FCN)-based encoder is deployed at the user side to map the received pilot signals into binary vectors, which serve as implicit channel features and are then fed back to the BS. For multi-user precoding, these binary vectors from all users are concatenated at the BS and processed by an FCN-based decoder to generate the beamforming matrix. 

Although the DL-based joint transceiver designs in \cite{Yu-TWC, GaoZhen_JSAC} demonstrate significant performance enhancement under limited pilot and feedback overhead, they suffer from poor scalability. Specifically, their proposed NNs should be redesigned and retrained whenever system parameters change. The scalability limitation stems from two factors: 1) the CSI extraction and feedback design compresses the extracted channel features into binary vectors of fixed length, making it incompatible with varying feedback capacities \cite{Bruno_GLOBECOM, LeeJungHoon_TWC, KB_TWC}; 2) the input and output dimensions of the multi-user precoding module are tied to the user count in the training dataset, rendering it ineffective when the number of users changes. Consequently, directly applying existing DL-based transceiver designs to practical FDD systems is challenging due to their limited adaptability to dynamic system configurations.

Several prior studies have partially addressed scalability concerns, focusing either exclusively on CSI extraction and feedback \cite{GuoJiajia_TWC, LinYuchien_TWC, Nerini_TWC} or on beamforming design \cite{Shenyifei_TWC, YangCY_TWC, Yu-JSAC, Rizzello_ICC}. For scalable CSI extraction and feedback with adaptive feedback capacities, \cite{GuoJiajia_TWC, LinYuchien_TWC} propose employing multiple encoders at the user side that can output vectors of different lengths based on the feedback capacity. \cite{Nerini_TWC} uses principal component analysis-based encoders to compress the channel into latent vectors with adaptive dimensionality. For scalable precoding design with varying numbers of users, graph neural networks (GNNs) have been employed in \cite{Shenyifei_TWC, YangCY_TWC, Yu-JSAC, Rizzello_ICC} to achieve robust multi-user precoding solutions, which learns the precoding policy from the graph representation of wireless networks and can be utilized in scenarios with an arbitrary number of users. However, these approaches individually address scalability either in the CSI extraction and feedback module or in the precoding module, without considering joint scalability. Moreover, directly integrating these scalable modules is impractical for two primary reasons. First, the output dimension of user-side encoders varies according to the feedback bit budget, causing dimension mismatches at the input of the multi-user precoding module. Second, existing GNN-based precoding designs assume either perfect CSI \cite{Shenyifei_TWC, YangCY_TWC} or perfect pilot signals \cite{Yu-JSAC, Rizzello_ICC} are available at the BS, which is impractical in low-overhead FDD systems. Therefore, it remains a major challenge to develop a unified and scalable DL-based framework capable of jointly optimizing pilot transmission, CSI extraction and feedback, and multi-user precoding, while providing improved generalization performance for cases with varying feedback capacities and numbers of users.

\subsection{Main Contributions}
This paper aims to design a scalable method for jointly optimizing pilot transmission, CSI extraction and feedback, and beamforming design for low-overhead communication in practical FDD massive MIMO systems using a data-driven approach. Specifically, our DL framework trains a dedicated NN to optimize the pilot signals. For CSI extraction and feedback, user-side encoders are used to extract channel features from the received pilot signals. These channel features are then quantized and fed back to the BS using a codebook-based vector quantization (VQ) technique. For multi-user precoding design, the BS concatenates the quantized channel features from all users and employs a BS-side NN to aggregate these features and derive the multi-user precoders. The key feature of the proposed transceiver design is its ability to utilize a single training phase to generate all the solutions required by the dynamic wireless setups. The main contributions of this work can be summarized as follows:
\begin{itemize}
    \item \textbf{Scalable design in terms of feedback codebook size}: 
    We propose a novel CSI feedback scheme based on the residual vector-quantized variational autoencoder (RVQ-VAE) to tackle the varying feedback capacities in FDD massive MIMO systems. The RVQ-VAE framework utilizes a hierarchical VQ codebook composed of multiple tiers of subcodebooks, enabling multi-resolution quantization that significantly enhances quantization efficiency compared to traditional methods. To ensure adaptability to dynamic feedback capacities, we introduce an efficient codebook size adjustment method that dynamically modifies the quantization bits required by the model. Furthermore, since the codeword dimension remains fixed, the reconstructed channel features at the BS always align with the input dimension of the multi-user precoding module, ensuring that the proposed transceiver is scalable to the feedback capacity.
    \item \textbf{Scalable design in terms of the number of users}: 
    We propose a feedback module sharing scheme combined with a novel edge graph attention network (EGAT) to address the challenge of varying user numbers. In this scheme, all users share the same feature extraction network and VQ codebook, eliminating the need to retrain the feedback module when new users join the system. For the multi-user precoding module, the EGAT employs a message-passing mechanism to learn the precoding policy from the channel features, making the precoding process independent of the number of users. Additionally, the attention mechanism in EGAT effectively captures inter-user interference, further enhancing the data transmission performance.
    \item \textbf{Progressive training strategy}: To fully exploit the representational power of RVQ, we propose a progressive training strategy that decomposes the training process into multiple stages. In the $n$-th stage, only the $n$-th tier of the codebook is updated, while the learnable parameters of other modules are fine-tuned, and the rest of the codebooks remain frozen. Compared with the end-to-end (E2E) training strategy, the proposed progressive training strategy can avoid overfitting for the proposed transceiver, leading to superior scalable performance.
    \item \textbf{Extensive numerical validation}: Numerical results based on a real-world channel dataset demonstrate that the proposed framework achieves sum-rate gains ranging from $18.9\%$ to $90.3\%$ over conventional ``estimate-then-feedback-then-beamform'' schemes and existing DL-based methods. Furthermore, extensive simulations validate the framework’s scalability concerning varying feedback capacities and user counts, while also highlighting its strong generalization capability to unseen environments.
\end{itemize}

\subsection{Organization}
The remainder of this paper is organized as follows. Section \ref{Sec. II} presents the system model. Section \ref{Sec. III} formulates the problem and introduces the DL-based approach used to solve it. Section \ref{Sec. IV} provides a detailed illustration of each module in the proposed NN framework. In Section \ref{Sec. V}, we analyze and address the scalability challenges associated with varying feedback capacities and user numbers. Section \ref{Sec. VI} evaluates the performance of the proposed scheme through extensive simulations. Finally, Section \ref{Sec. VII} concludes the paper.

\textit{Notation}: Column vectors and matrices are denoted by boldfaced lowercase and uppercase letters, e.g., $\xb$ and $\Xb$. $\Zbb^+$ and $\Rbb^+$ denote the set for positive integers and positive real numbers, respectively $\Rbb^{n \times n}$ and $\Cbb^{n \times n}$ stand for the sets of $n$-dimensional real and complex matrices, respectively. The superscripts $(\cdot)^T$ and $(\cdot)^H$ describe the transpose and conjugate transpose operations, respectively. $\| \xb \|_2^2$ and $\| \Xb \|_F^2$ denote squares of the $l_2$-norm and Frobenius norm of vector $\xb$ and matrix $\Xb$, respectively. The statistical expectation operator is represented by $\Ebb[\cdot]$.

\section{System Model} \label{Sec. II}
We consider a downlink FDD multi-user massive MIMO system with a block-fading channel model, where a BS with $M$ antennas serves $K$ single-antenna users. Since the uplink-downlink channel reciprocity does not hold in the FDD system, the BS should broadcast a pilot sequence of length $L$ (in terms of samples) to obtain some information about the downlink channels for precoding design. Define $\tilde{\Xb} \triangleq [\tilde{\xb}_1, \tilde{\xb}_2, \cdots, \tilde{\xb}_L] \in \Cbb^{M \x L}$ as the pilot matrix, where $\tilde{\xb}_l$ denotes the $l$-th pilot samples transmitted by all the BS antennas and satisfies the power constraint $\| \tilde{\xb}_l \|_2^2 \leq P$, $\forall l \in \{1,\cdots, L\}$. The received signal of the $k$-th user at the pilot transmission phase is given as
\begin{equation} \label{Rx pilot}
    \tilde{\yb}_k^H = \hb_k^H \tilde{\Xb} + \tilde{\nb}_k^H, \quad \forall k,
\end{equation}
where $\hb_k \in \Cbb^{M \x 1}$, $\forall k \in \{1, 2, \cdots, K\}$, is the downlink channel vector between the BS and user $k$, and $\tilde{\nb}_k \sim \mathcal{CN}(\mathbf{0}, \sigma^2\Ib_L)$ is the additive white Gaussian noise (AWGN) at user $k$ at the pilot transmission phase, with $\sigma^2$ specifying the noise power and $\Ib_L$ being the identity matrix of size $L$.  

Because it is practically impossible to accurately estimate and feed back user channels in downlink massive MIMO communication, we propose to extract the channel features that are crucial for subsequent BS precoding design from the pilot signals and feed back these feature vectors using a low-overhead VQ-based mechanism. Specifically, upon receiving the pilot signal given in \eqref{Rx pilot}, the $k$-th user first extracts the channel feature and expresses it using $D$ symbols, i.e.,
\begin{equation}  \label{F}
	\vb_k = \mathcal{F} \left( \tilde{\yb}_k \right) \in \Rbb^{D \x 1}, \quad \forall k,
\end{equation}
where $\mathcal{F}:\Cbb^{L \x 1} \rightarrow \Rbb^{D \x 1}$ is the common channel feature extraction function shared for all users. Then, each user $k$ quantizes its channel feature vector with $B$ quantization bits using a low-overhead VQ-based method, which can be expressed as
\begin{equation} \label{Qe}
	\bb_k = \mathcal{Q}_{e_B}\left( \vb_k \right), \quad \forall k,
\end{equation}
where $\bb_k = [b_{k,1}, b_{k,2}, \cdots,b_{k, B}]^T \in \{ 0, 1 \}^B$ is a binary vector, and $\mathcal{Q}_{e_B}: \Rbb^{D \x 1} \rightarrow \{0,1\}^B$ specifies the shared quantization function for all users. Subsequently, each user $k$ feeds back the binary vector $\bb_k$ to the BS. After collecting the binary vectors from all users, the BS applies a common dequantization function $\mathcal{Q}_{d_B}: \{0,1\}^B \rightarrow \Rbb^{D \x 1}$ to reconstruct the channel feature vector $\vb_k$ of each user $k$ as
\begin{equation}
    \hat{\vb}_k(B) = \mathcal{Q}_{d_B} \left( \bb_k \right), \quad \forall k,
\end{equation}
where $\hat{\vb}_k(B)$ denotes the dequantized result of $\vb_k$ under the feedback capacity of $B$. Thus, the entire CSI extraction and feedback process can be represented by a composite feedback module $\mathcal{V}_B:\Cbb^{L\times 1}\rightarrow\Rbb^{D\times 1}$, which is given by
\begin{equation} \label{eqn: V_cal}
    \hat{\vb}_k(B) \triangleq \mathcal{V}_B(\tilde{\yb}_k) = \mathcal{Q}_{d_B}( \mathcal{Q}_{e_B}( \mathcal{F}( \tilde{\yb}_k ) ) ), \quad \forall k.
\end{equation}

The BS then aggregates all the dequantized channel features into a feature matrix $\hat{\Vb}_K(B) \triangleq [\hat{\vb}_1(B),\hat{\vb}_2(B),$ $\cdots, \hat{\vb}_K(B)] \in \Rbb^{D \x K}$. Finally, the BS designs the downlink multi-user precoding matrix $\Wb_K(B) \triangleq [\wb_1(B), \wb_2(B), \cdots, \wb_K(B)] \in \Cbb^{M \x K}$, where $\wb_k(B)$ denotes the precoding vector to user $k$ under the feedback bit budget of $B$, $\forall k$. This process is described as follows

\begin{equation} \label{P and Qd}
    \Wb_K(B) = \mathcal{W}_K \left( \hat{\Vb}_K(B), P \right),
\end{equation}
where $\mathcal{W}_K: \Rbb^{D \x K} \x \Rbb^+ \rightarrow \Cbb^{M \x K}$ is the function mapping the recovered channel features of $K$ users and the available transmission power to the multi-user precoding matrix.

Define the data signal transmitted by the BS as $\xb(B) = \sum_{k=1}^{K} \wb_k(B) s_k$, where $s_k \sim \mathcal{CN}(0,1)$ is the data symbol of user $k$. Then, the received signal at user $k$ in the data transmission phase given the feedback capacity of $B$ can be written as 
\begin{equation} \label{Rx signal}
    y_k(B) = \hb_k^H \wb_k(B) s_k + \sum_{j \neq k}\hb_k^H \wb_j(B) s_j + n_k, \quad \forall k,
\end{equation}
where $n_k \sim \mathcal{CN}(0, \sigma^2)$ is the AWGN at user $k$ at the data transmission phase. Then, the achievable rate of user $k$, denoted as $R_k(\hb_k, \Wb_K(B))$,  can be defined as
\begin{align} \label{eqn: Rk}
    R_k(\hb_k, \Wb_K(B)) &= \notag \\
    &\log_2 \left( 1 + \dfrac{ |\hb_k^H \wb_k(B)|^2 }{ \sum_{j \neq k} | \hb_k^H \wb_j(B) |^2 + \sigma^2 } 
    \right), \forall k.
\end{align}

\section{Problem Description and Scalable DL Solutions} \label{Sec. III}

\subsection{Problem Description}
In practice, the number of users, i.e., $K$, and the user channels, i.e., $\hb_k$'s, may change over time. Correspondingly, each user may also change its number of bits for quantization, i.e., $B$, over time. Suppose the distributions of $B$, $K$, and $\hb_k$ can be known based on historical data. Moreover, define $\mathcal{K}$ and $\mathcal{B}$ as the sets containing all the possible values of $K$ and $B$ under their distributions, respectively. In this paper, we aim to design the pilot matrix $\tilde{\Xb}$, the implicit channel feature feedback module $\mathcal{V}_B(\cdot)$, $ \forall B \in \mathcal{B} $, and the precoding scheme $\mathcal{W}_K(\cdot)$, $\forall K \in \mathcal{K}$, which is optimal to maximize the user's average sum rate in the long term. The corresponding optimization problem can be formulated as

\begin{equation} \label{Opt problem}
\begin{array}{ll}
    \mathop{\max}\limits_{\substack{\tilde{\Xb}, \mathcal{F}(\cdot), \\ \{ \mathcal{V}_{B}(\cdot) \}_{ B \in \mathcal{B}}, \\ \{ \mathcal{W}_K(\cdot) \}_{K \in \mathcal{K}} }} & \Ebb_{B, K, \hb_k} \left[ \sum\limits_{k=1}^K R_k(\hb_k, \Wb_K(B)) \right] \\

    \text{s.t.} & \Wb_K(B) = \mathcal{W}_K \left( \hat{\Vb}_K(B), P \right),~~\forall B, K,\\
    & \hat{\Vb}_K(B) = [\mathcal{V}_B \left( \tilde{\yb}_1 \right), \cdots, \mathcal{V}_B \left( \tilde{\yb}_K \right)],~~\forall B, K,\\
      & \operatorname{Tr}\left( \Wb_K(B) (\Wb_K(B))^H \right) \leq P,~~\forall B, K, \\
      & \| \tilde{\xb}_l \|_2^2 \leq P, \quad \forall l. \\
\end{array}
\end{equation}
Note that problem \eqref{Opt problem} involves both variable and function optimization. The non-convex nature of the sum-rate maximization objective and the implicit expressions of the mapping functions make it highly challenging to solve using traditional optimization methods. To address this difficulty, we adopt the DL-based approach, which leverages NNs to model and optimize these functions and variables.

\subsection{DL-Based Framework and Scalable Solutions}
Existing studies \cite{Yu-TWC, GaoZhen_JSAC} have proposed DL-based transceiver designs to address a special case of problem \eqref{Opt problem}, where the number of feedback bits and users are both fixed, i.e., $\mathcal{B} = \{B_0\}$ and $\mathcal{K} = \{K_0\}$. These approaches rely on FCNs to model key transceiver components, including channel feedback and multi-user precoding, in which the input/output dimensions of these NNs are fixed and predetermined by $(B_0, K_0)$ used during training. As a result, these DL-based methods lack scalability, meaning that their NNs can only work under the system configuration $(B_0, K_0)$ and become inapplicable when the system parameter $(B, K)$ changes during deployment. As such, to solve problem \eqref{Opt problem} with $|\mathcal{B}| > 1$ and $|\mathcal{K}| > 1$, we need to train and store an individual NN model for each possible $(B, K)$ pair, leading to significant computational and storage overhead in practical deployments.

To address this limitation, this paper aims to design a scalable DL-based transceiver, in which a single NN model, trained only once using data sampled from the distribution of $B$, $K$, and $\hb_k$, can be deployed to have consistent performance across different $(B, K)$ configurations. However, designing such a scalable framework presents two critical challenges. First, variations in $B$ directly affect the accuracy of the channel feature matrix $\hat{\Vb}_K(B)$, which serves as the input to the precoding network. Thus, changing $B$ impacts the design of both feedback and precoding functions. Second, variations in $K$ require the precoding network to dynamically adapt its output dimension to produce appropriate beamformers for all active users. Therefore, the key technical challenge lies in designing a single unified NN architecture capable of adaptively realizing channel feedback $\mathcal{V}_{B}(\cdot)$ across varying values of $B$, while simultaneously adjusting the precoding function $\mathcal{W}_{K}(\cdot)$ for varying $K$. To address these challenges, we propose a novel framework that jointly incorporates a multi-resolution feedback scheme with a scalable multi-user precoding network. In the next section, we provide a detailed explanation of each module and illustrate how this framework enhances both scalability and overall system performance.

\section{Proposed DL-Based Transceiver Design} \label{Sec. IV}
\begin{figure*}[!t]
\centerline{\includegraphics[width=0.9\textwidth]{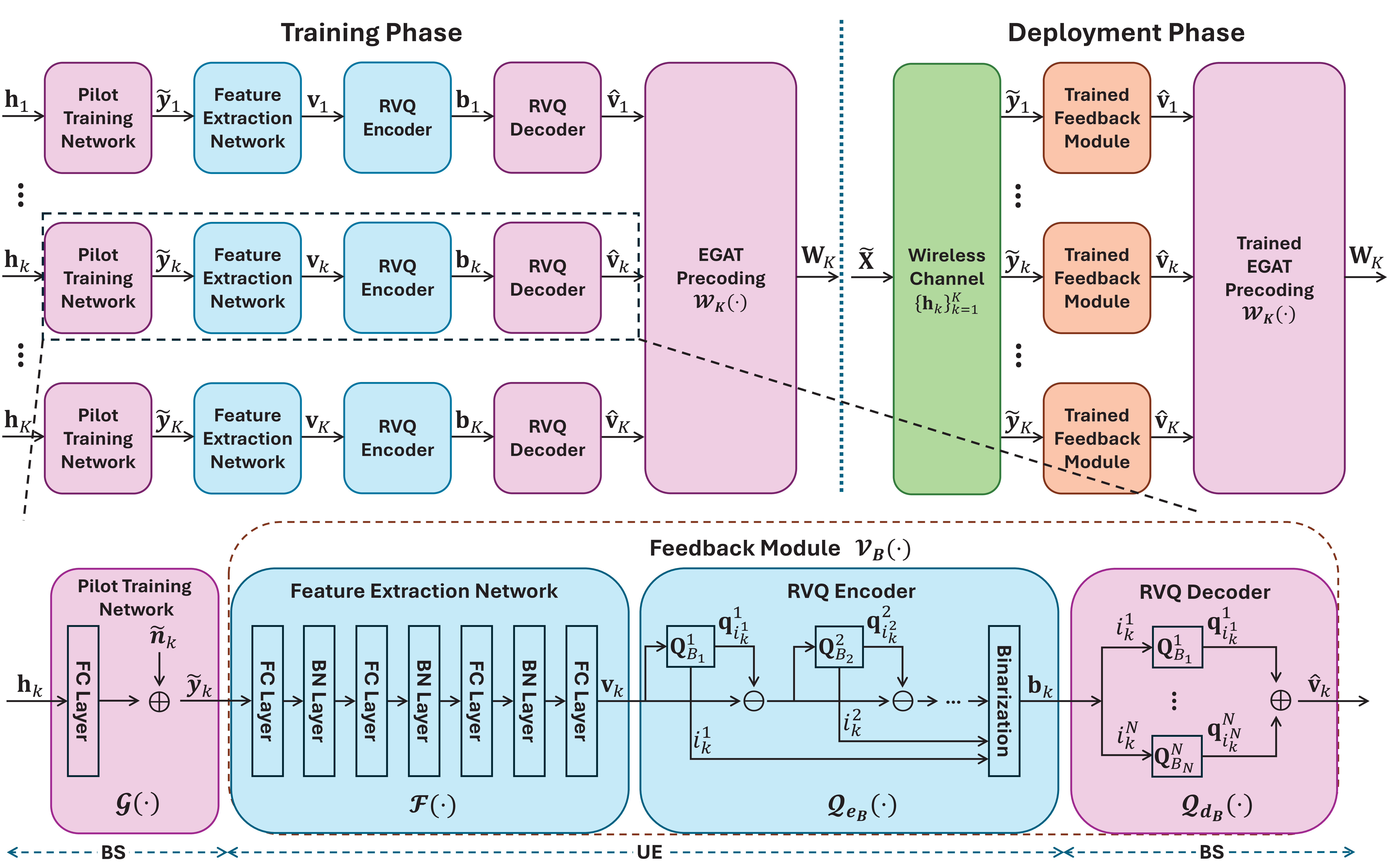}}
\caption{Block diagram of the proposed DL-based framework, where the dependencies on $B$ for $\hat{\vb}_k$ and $\Wb_K$ are omitted for the sake of notation simplicity.}
\label{Fig: System diagram}
\end{figure*}

This section presents the proposed DL-based approach for designing a scalable transceiver in the considered system. The architecture of the transceiver in the training and deployment phases is illustrated in Fig. \ref{Fig: System diagram}. During the training phase, three key components are trained: a pilot training network for improved pilot transmission, a feedback module for flexible channel feedback, and an EGAT-based multi-user precoding network for scalable beamforming design. Specifically, the feedback module employs RVQ with a hierarchical quantization codebook, which can dynamically adjust the codebook size to achieve multi-resolution quantization. The EGAT-based precoding network relies on the GNN architecture that can learn the multi-user communication topology to adaptively design beamformers for different numbers of users. In this structure, each user node processes its feature through a shared function, exchanges information with neighboring nodes, and performs local feature updates, thus enabling the precoding process inherently independent of $K$.

In the deployment phase, the BS first transmits the trained pilot matrix $\tilde{\Xb}$ to users for channel acquisition. Each user $k$ then applies the trained feature extractor $\mathcal{F}(\cdot)$ and quantizer $\mathcal{Q}_{e_B}(\cdot)$ to generate feedback information for the BS. Upon receiving the feedback, the BS reconstructs the users’ channel features using the trained dequantizer $\mathcal{Q}_{d_B}(\cdot)$ and computes the optimized beamformer using the trained precoding network $\mathcal{W}_K(\cdot)$. Note that each transceiver function is parameterized by an NN and generally denoted by $\mathcal{F}(\cdot;\triangleright)$, where $\cdot$ denotes the input and $\triangleright$ denotes the NN parameters. The following subsections provide a detailed design of each of these three components.
%\begin{remark}
%It is worth noting that in \cite{Yu-TWC, GaoZhen_JSAC}, different users are assigned different feature extractors, quantizers, and dequantizers during the training phase. This leads to the scalability issue in the deployment phase, as their pre-trained models fail to provide dedicated functions for newly added users. To overcome this issue, we propose to adopt a common feature extractor, quantizer, and dequantizer for all users. As will be shown in Section \ref{Sec. V-B}, such a design can effectively decouple the feedback mechanism from the number of users, facilitating the scalable transceiver design.
%\end{remark}
%\begin{remark}
%It is worth noting that in \cite{Yu-TWC, GaoZhen_JSAC}, different users are assigned different feature extractors, quantizers, and dequantizers during the training phase. This leads to the scalability issue in the deployment phase, as their pre-trained models fail to provide dedicated functions for newly added users. To overcome this issue, we propose to adopt a common feature extractor, quantizer, and dequantizer for all users. As will be shown in Section \ref{Sec. V-B}, such a design can effectively decouple the feedback mechanism from the number of users, facilitating the scalable transceiver design.
%\end{remark}

\subsection{Pilot Training Network}
The pilot training network aims to generate the optimal pilot sequence that can effectively capture useful channel information for the multi-user precoding design. Notably, the part of noiseless pilot observation in \eqref{Rx pilot} resembles the signal processing mechanism in a single-layer FCN. Therefore, the pilot transmission procedure can be modeled by a single-layer FCN in the training phase, where the channel of user $k$, i.e., $\hb_{k}$, serves as input and the learnable pilot matrix, i.e., $\tilde{\Xb}$, is considered as the matrix of network weight parameters $\forall k$. By adding the background noise at the output of the single-layer FCN, the received pilot signals of all users can be finally obtained. Thus, we can express the pilot training network as

\begin{equation}
    \tilde{\yb}_k = \mathcal{G}\left( \hb_k; \tilde{\Xb} \right), \quad \forall k,
\end{equation}
To ensure the weight matrix $\tilde{\Xb}$ satisfies the pilot transmission power constraint, we always normalize each column of $\tilde{\Xb}$ by $\tilde{\xb}_l = \sqrt{P} \tilde{\xb}_l / \| \tilde{\xb}_l \|_2$, $\forall l$.

\subsection{Feedback Module} \label{Sec. IV-B}
The feedback module in the proposed DL-based framework is responsible for achieving both functions of extraction and quantization/dequantization for the channel features embedded in the pilot signals. To accomplish these tasks, three sub-networks for feature extraction, quantization, and dequantization are included in the feedback module, as illustrated in Fig. \ref{Fig: System diagram}. For feature extraction, we adopt a multi-layer FCN to process the received pilot signals, which can be expressed as
\begin{equation}
\vb_k = \mathcal{F}(\tilde{\yb}_k;\mathbf{\Theta}_{F}) \in \Rbb^{D \x 1}, \quad \forall k.
\end{equation}
where $\mathbf{\Theta}_{F}$ denotes the weight matrix for the feature extraction network. For quantization and dequantization, we leverage the VQ-VAE architecture \cite{Taomx-Lett, VQVAE}, which integrates the VQ method with DL techniques. In this setup, the VQ encoder quantizes the extracted feature vector $\vb_k$ at each user $k$, while the VQ decoder reconstructs $\vb_k$ at the BS. Meanwhile, the latent variable in the form of a $B$-bit binary vector is considered as the feedback information from the user to the BS. Note that VQ-VAE provides an explicit discretization of the input space, and its VQ codebook is treated as a learnable parameter, allowing it to be optimized during training for improved quantization efficiency.

While VQ-VAE offers an effective approach to CSI feedback, its training and deployment pose two key challenges. First, the size of the learnable codebook grows exponentially with increasing feedback capacity, leading to substantial storage demands and computational complexity. Second, the absence of a well-defined gradient in the codeword selection process makes traditional back-propagation unsuitable for updating the parameters of VQ-VAE. To overcome these challenges, we propose two dedicated solutions to separately address these challenges.

To address the first challenge, we propose an RVQ-VAE for CSI feedback, where the novel RVQ method significantly reduces codebook size while preserving quantization performance. Unlike conventional approaches that directly select the closest codeword from a single predefined codebook, RVQ applies a hierarchical quantization technique that divides the process into multiple stages, with each stage having its own codebook. Initially, the input vector is quantized using the first stage codebook, and the residual is calculated. This residual is then passed to the second stage for further quantization, and the process continues until all stages are completed. The final quantized result is obtained by summing the quantized outputs from all stages.

Under the proposed RVQ-VAE architecture, each user $k$ is assigned an $N$-tier codebook $\Qb_B \triangleq \{\Qb_{B_1}^1, \Qb_{B_2}^2, \cdots, \Qb_{B_N}^N\}$ for quantizing the channel feature vector $\vb_k$. The quantization process is decomposed into $N$ stages, where each stage $n$ is associated with a subcodebook $\Qb_{B_n}^n \triangleq [\qb_{1}^n,\qb_{2}^n,\cdots,\qb_{2^{B_n}}^n] \in \Rbb^{D \x 2^{B_n}}$. Here, $B_n$ is the number of quantization bits allocated to stage $n$ satisfying $\sum_{n=1}^{N} B_n = B$. Let $\mathcal{I}_n \triangleq \{ 1,2,\cdots, 2^B_n \}$ denote the set of all codeword indices in $\Qb^n$. For each user $k$, we define $\rb^n_{k}$ as the residual vector between the input and the quantized output at stage $n$, which is initialized as $\rb^0_k = \vb_k$. At the $n$-th quantization stage, the input is the residual vector from the previous stage, i.e., $\rb^{n-1}_{k}$. The RVQ encoder first identifies the codeword $\qb_{i_k^n}^n$ that is closest to $\rb^{n-1}_{k}$ in terms of Euclidean distance and then encodes its index $i_k^n$ into a binary vector $\bb_k^n \in \{0,1\}^B_n$ for feedback, which can be expressed as
\begin{subequations} \label{eqn: i_k^n and A}
\begin{align}
    i_k^n &= \mathop{\arg \min}\limits_{j \in \mathcal{I}_n } \| \qb^n_{j} - \rb^{n-1}_{k} \|_2^2, \quad \forall k,n, \\
    \bb_k^n &= \mathcal{A} \left( i_k^n, B_n\right) \in \left\{ 0, 1\right\}^{B_n}, \quad \forall k,n,
\end{align}
\end{subequations}
where $\mathcal{A}( i_k^n, B_n)$ is a function that converts the index $i_k^n$ into a binary vector of length $B_n$, $\forall k,n$. Based on the selected codeword, the residual vector is then updated as 
\begin{equation} \label{eqn: Update r_k^n}
    \rb^n_{k} = \rb^{n-1}_k - \qb_{i_k^n}^n, \quad \forall k,n.
\end{equation}
This updated residual vector serves as the input to the $(n+1)$-th quantization stage. After completing $N$ stages, the final feedback information for user $k$ is obtained by concatenating the binary vectors from all stages as $\bb_k = [(\bb_k^1)^T,\cdots,(\bb_k^N)^T]^T \in \{0,1\}^B$. The entire quantization and binarization process performed by the RVQ encoder can be summarized as
\begin{equation}
    \bb_k = \mathcal{Q}_{e_B} \left( \vb_k; \Qb_B \right), \quad \forall k
\end{equation}

Upon receiving the feedback information $\bb_k$ from each user $k$, the BS utilizes the corresponding RVQ decoder to reconstruct the channel feature vector through the dequantization process, which is also carried out in $N$ stages. For each stage $n$, the RVQ decoder for each user $k$ retrieves the $i_k^n$-th codeword $\qb_{i_k^n}^n$ from the trained codebook $\Qb_{B_n}^n$. The final reconstructed channel feature vector is then obtained by summing the selected codewords across all stages, which is given by
\begin{equation} \label{eqn: RVQ v_k_hat}
    \hat{\vb}_k(B) = \sum_{n=1}^{N} \qb_{i_k^n}^n, \quad \forall k.
\end{equation}
The complete dequantization process performed by the RVQ decoder is expressed as
\begin{equation} \label{Q_d}
    \hat{\vb}_k(B) = \mathcal{Q}_{d_B}\left( \bb_k; \Qb_{B} \right), \quad \forall k.
\end{equation}

Compared to the conventional VQ method that uses a single-tier codebook, RVQ offers a more flexible solution to control the complexity by decomposing the quantization process into $N$ stages using an $N$-tier codebook. In the conventional VQ method, the number of required codewords grows exponentially as $2^B$, leading to high storage and computational complexity. In contrast, RVQ requires a total number of codewords equal to $\sum_{n=1}^N 2^{B_n} \ll 2^B$, allowing for a more efficient codebook storage and quantization procedure. In this work, we adopt the equal bit allocation strategy for RVQ codebook design, i.e., $B_n = \bar{B} \triangleq B / N$, $\forall n$, which is validated to achieve satisfactory performance in a wide range of feedback capacities by simulation results.
\begin{remark}
The hierarchical structure of the codebook design in the proposed RVQ-VAE model also facilitates scalable design for varying feedback capacities. If the feedback capacity is smaller than the one used during training, we can directly discard the later VQ layers to reduce the number of quantization bits. On the contrary, if the feedback capacity exceeds that in the training phase, we can add more VQ layers without modifying the previously trained model, requiring only a small additional training overhead. Further details on the scalable design of the RVQ-VAE are provided in Section \ref{Sec. V-A}.
\end{remark}

For the second challenge, we propose to employ the noise substitution vector quantization (NSVQ) technique \cite{NSVQ}. In particular, NSVQ approximates the quantization process by introducing a noise vector into the original vector. This noise vector is crafted to mimic the distribution of the quantization error, thereby facilitating gradient-based optimization. For the proposed RVQ-based quantization scheme, recall that $\vb_k$ is the input to the first VQ stage and $\hat{\vb}_k(B)$ is the final quantized result after $N$-stage quantization. The NSVQ method aims to approximate the quantized output of $\vb_k$ as
\begin{equation} \label{approx. vq}
    \hat{\vb}_{k}(B) = \vb_k + \dfrac{\| \vb_k - \hat{\vb}_{k}(B) \|_2}{\| \ub \|_2} \ub, \quad \forall k,
\end{equation}
where $\ub \sim \mathcal{CN}(\mathbf{0}, \Ib)$ is a random noise vector. This approximation allows the gradients of the codeword selection process to be computed and propagated to the preceding layers during the back-propagation process. Compared with the conventional straight-through estimator \cite{Yu-TWC} that approximates the gradient of VQ using a smooth differentiable function, NSVQ can offer superior performance in accuracy and convergence speed \cite{NSVQ}.

\subsection{Edge Graph Attention Network for Multi-User Precoding} \label{Sec. IV-C}
\begin{figure}[!t]
\centerline{\includegraphics[width=0.4\textwidth]{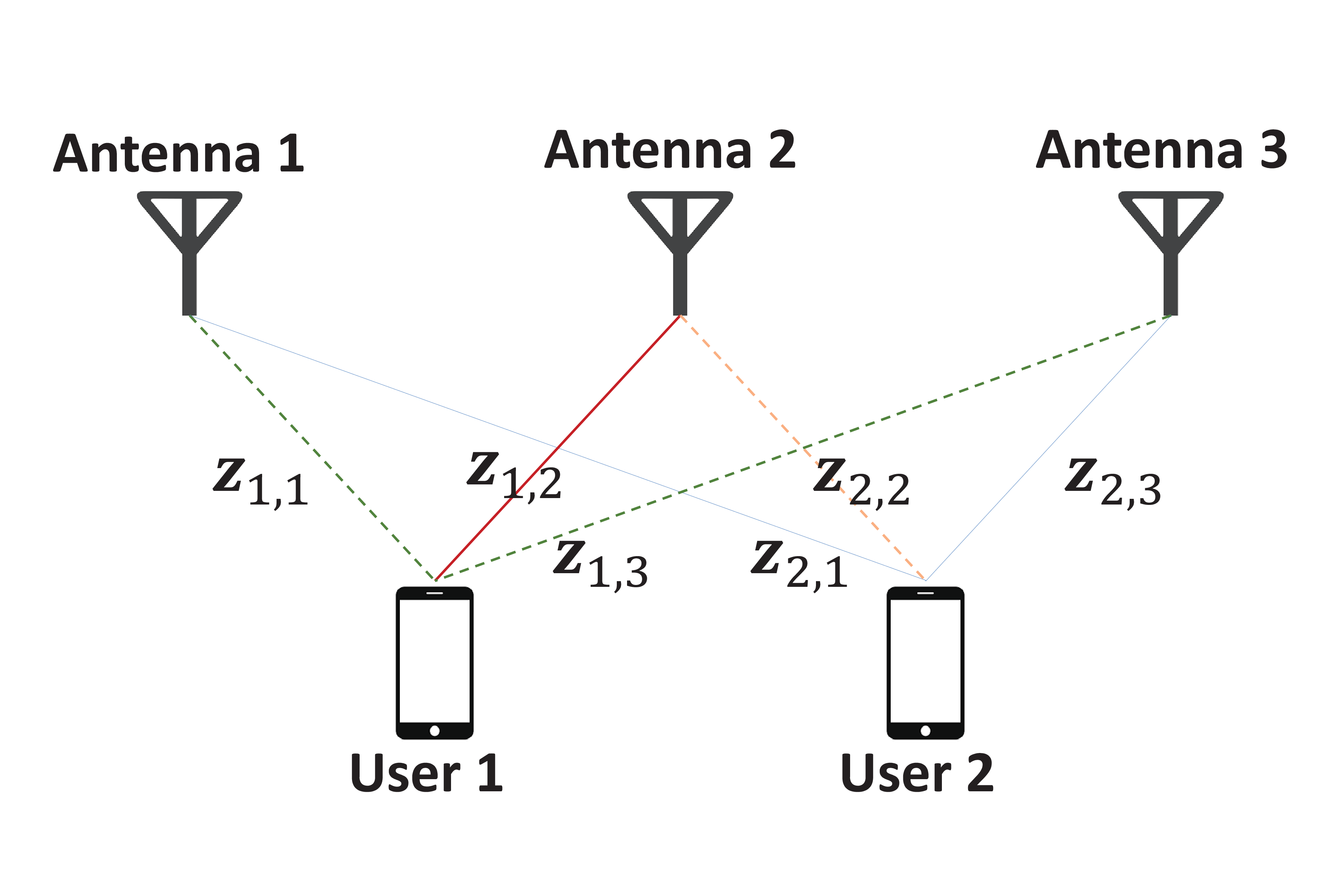}}
\caption{Edge representation of an multi-user MIMO system with $M=3$ and $K=2$. For edge $(1,2)$, the neighboring edges include orange dashed lines (connections from Antenna $2$ to other users) and green dashed lines (connections from User $1$ to other antennas).}
\label{Fig: Edge Representation}
\end{figure}

\begin{figure*}[!t]
\centerline{\includegraphics[width=0.9\textwidth]{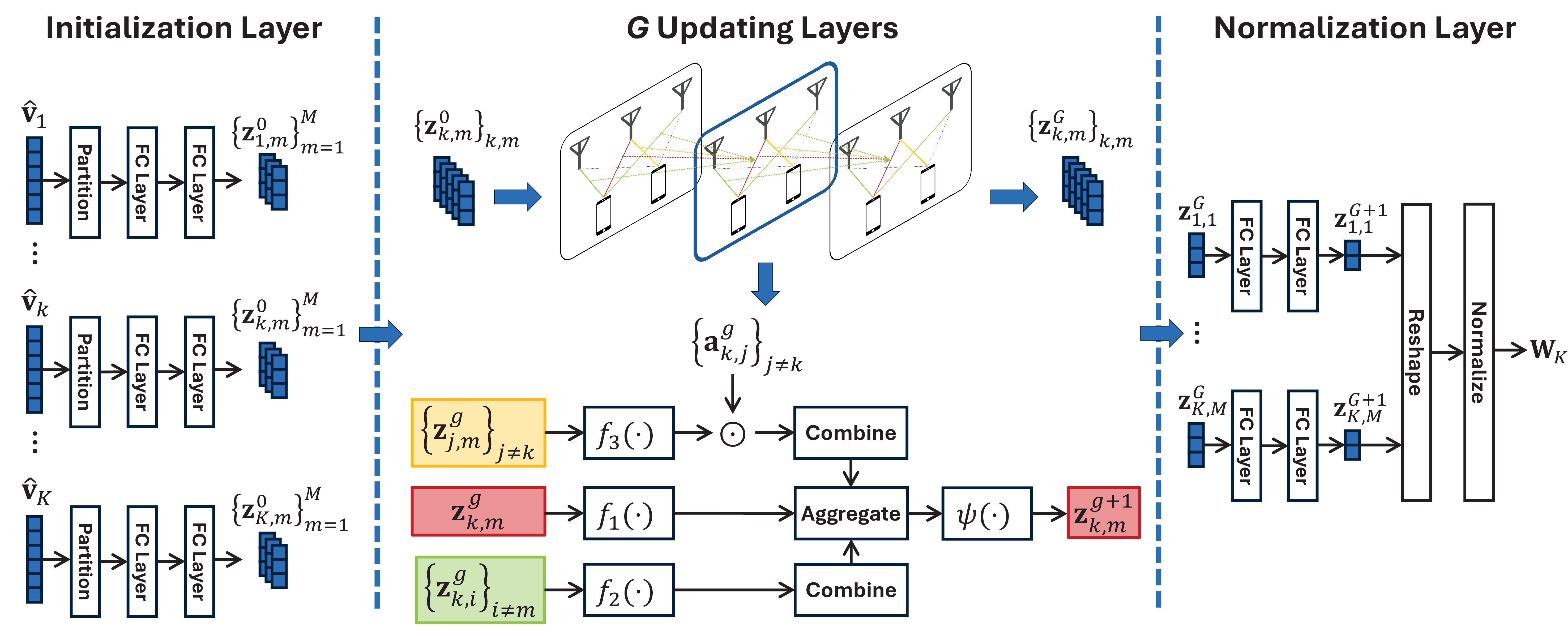}}
\caption{Structure of the proposed EGAT precoding design.}
\label{Fig: EGAT}
\end{figure*}

After reconstructing the channel feature vectors, we propose utilizing an EGAT that integrates the attention mechanism and edge-based GNN for the multi-user precoding design. We first construct the graphical representation of the multi-user MIMO system using a bipartite graph with $K$ user nodes and $M$ antenna nodes, as shown in Fig. \ref{Fig: Edge Representation}. Here, each user node $k$ is linked to each antenna node $m$ through an edge denoted as $(k,m)$, $\forall k,m$. In particular, each edge $(k,m)$ is associated with a state vector, $\zb_{k,m}$, $\forall k,m$, which will be iteratively updated within the EGAT. Upon the built graph representation, we then design an EGAT consisting of an initialization layer, followed by $G$ updating layers, and a final normalization layer, as illustrated in Fig. \ref{Fig: EGAT}.

This work highlights three key advantages of employing EGAT for precoding design over conventional NN-based methods. First, unlike the FCN-based precoding networks in \cite{Yu-TWC, GaoZhen_JSAC}, EGAT is built on GNNs, which offers reduced model complexity and improved scalability. In terms of model complexity, GNNs employ shared state-updating functions across nodes and edges, leading to fewer parameters than FCNs, which require distinct weights for each connection. Additionally, EGAT inherently accounts for two-dimensional permutation equivariance (2D-PE) properties in sum-rate maximization problem, ensuring that the order of users and transmit antennas does not affect the sum-rate but the order of the optimal precoding vectors. This enables EGAT to handle permuted datasets without additional training, whereas FCNs require a large model and augmented datasets containing all possible permutations. From a scalability standpoint, FCN-based approaches have fixed input/output dimensions tied to the number of users in the training dataset, requiring redesigning and retraining whenever the user count changes. In contrast, the state-updating functions and message-passing mechanism make the precoding process of EGAT independent of the number of users, allowing it to adapt to scenarios with arbitrary user numbers.

Second, compared with existing GNN-based precoding networks \cite{Shenyifei_TWC, Yu-JSAC} that only update node representations and thus can capture only PE with respect to user indices, the proposed EGAT updates edge representations to fully capture the 2D-PE properties inherent in \eqref{Opt problem} in terms of indices for both users and antennas. By capturing the 2D-PE properties, the network complexity can be further reduced and there is no need to utilize an augmented dataset containing all possible channel permutations.
 
Third, compared with the conventional message-passing GNNs that rely solely on FCNs to process state representations, the proposed EGAT integrates an attention mechanism to dynamically measure the importance of each neighboring state based on its relevance to the target state. This attention mechanism allows the network to effectively model and capture the interference between users, leading to improved precoding decisions and enhanced data transmission performance \cite{Yangcy-WCNC}.

In the following, we give the details of the proposed EGAT for multi-user precoding design.

\subsubsection{Initialization Layer} The initialization layer is constructed using a two-layer FCN, denoted as $f_0(\cdot)$, which generates the initial state vectors for all edges based on the quantized channel feature vectors of each user, $\{\hat{\vb}_{k}\}_{k=1}^{K}$.\footnote{For simplicity, the dependence on the parameter $B$ is omitted in this subsection.} Specifically, $\hat{\vb}_{k}$ is first partitioned as $\hat{\vb}_{k} = [\hat{\vb}_{k,1}^T,\hat{\vb}_{k,2}^T,\cdots,\hat{\vb}_{k, M}^T]^T$, where $\hat{\vb}_{k,m} \in \mathbb{R}^{D/M \x 1}$ represents the initial feature information for the edge connecting user node $k$ and antenna node $m$. The initial state vector for each edge $(k,m)$, denoted as $\zb_{k,m}^0$, is then computed as $\zb_{k,m}^0 = f_0\left( \hat{\vb}_{k,m} \right)$. These initialized state vectors are subsequently passed through $G$ updating layers to refine and update the edge states.

\subsubsection{Updating Layers}
The update of the state vector $\zb_{k,m}^g$ of edge $(k,m)$ in the $g$-th updating layer, $\forall g \in \{0, 1, \cdots, G-1\}$, is based on a combination of the processed result of $\zb_{k,m}^g$ and the aggregated information from neighboring edges. Specifically, the state of edge $(k,m)$ after passing through the $g$-th updating layer, $\mathbf{z}_{k,m}^{g+1}$, can be expressed as
\begin{align} \label{update nk}
    \zb_{k,m}^{g+1} = \psi &\Big( f_1^{g} \big( \zb_{k,m}^{g} \big) + \dfrac{\alpha}{M}\sum_{i=1,i\neq m}^{M} f_2^g \big( \zb_{k,i}^g \big)\notag  \\  &\qquad+ \beta \sum_{j=1,j\neq k}^{K} \mathbf{a}_{k,j}^g \odot f_3^g \big( \zb_{j,m}^g \big) \Big), ~ \forall k,m,
\end{align}
where 
\begin{equation}
\mathbf{a}_{k,j}^{g} = \dfrac{1}{M} \sum_{m=1}^{M} f_{4}^{g}(\zb_{k,m}^{g}) \odot f_{5}^{g}(\zb_{j,m}^{g}).
\end{equation}
In \eqref{update nk}, $f_i^g(\cdot)$, $\forall i \in \{1,2,\cdots,5\}$, $\forall g$, is the linear processing function implemented via single-layer FCN, and $\psi(\cdot)$ is the element-wise activation function. $\odot$ denotes the Hadamard product. $\alpha$ and $\beta$ are hyperparameters that can balance the contribution of each component to get better convergence performance. In this equation, the first term processes the current edge state, while the second term aggregates the states of neighboring edges from other antenna nodes. The third term aggregates information from other user nodes, which reflects the influence (interference) of other users on user $k$. To capture this interference, we introduce an attention coefficient vector $\ab_{k,j}^g$, which dynamically adjusts the weight of the influence of user $j$ on user $k$. This attention mechanism helps in modeling the interference between users, thereby improving the precoding performance \cite{Yangcy-WCNC}.

\subsubsection{Normalization Layer}
After passing through $G$ updating layers, the state vectors $\{\zb_{k,m}^G\}_{\forall k, m}$ are fed into a normalization layer to generate the precoding matrix $\Wb_K \in \Cbb^{M \x K}$. Specifically, $\zb_{k,m}^G$ is passed through a two-layer FCN, denoted as $f_N(\cdot)$, which generates the real and imaginary components of the $(k,m)$-th complex entry of $\Wb_K$, $\forall k, m$. This transformation results in $\zb_{k,m}^{G+1} = f_N(\zb_{k,m}^G) \in \Rbb^{2}$. The $(k,m)$-th entry of $\Wb_K$, $w_{k,m}$, is then constructed by reshaping $\zb_{k}^{G+1}$ as
\begin{equation} \label{w_k}
    w_{k,m} = \zb_{k,m}^{G+1}\left( 1 \right) + j\zb_{k,m}^{G+1}\left( 2 \right), \quad \forall k,
\end{equation}
where $\zb(i)$, $\forall i \in \{1,2\}$, denotes the $i$-th entry of $\zb$. Finally, the precoding matrix $\Wb_K$ is normalized to satisfy the power constraint as $\Wb_K = \sqrt{P} \Wb_K / \| \Wb_K \|_F$, where $\| \cdot \|_F$ represents the Frobenius norm. One should mention that the state-updating functions of the proposed EGAT are shared across different user nodes. Such parameter sharing enables the network to generalize the precoding solution to an arbitrary number of users. Thus, the overall precoding process can be expressed as
\begin{equation} \label{GNN}
    \Wb_K = \mathcal{W}_K\left( \hat{\Vb}_K, P ; \mathbf{\Theta}_W \right),
\end{equation}
where $\mathbf{\Theta}_W$ represents all the parameters of the EGAT layers. Note that the transmit power $P$ is also input to the NN, making the proposed DL scheme also scalable to arbitrary $P$.

\subsection{Training Policy} \label{Sec. IV-D}
We propose to jointly optimize the parameters of the pilot network, feedback module, and the proposed EGAT. The total loss function of the proposed network is defined as
\begin{equation} \label{total loss}
    \mathcal{L} = \mathcal{L}_R + \lambda \mathcal{L}_Q,
\end{equation}
where $\mathcal{L}_R \triangleq -\sum_{k=1}^K R_k $ is the negative of the sum-rate with $R_k$ defined in \eqref{eqn: Rk}. $\mathcal{L}_Q$ is the quantization loss. Specifically, the mean squared error (MSE) between the true channel feature vectors $\{\vb_k\}_{k=1}^K$ and the corresponding quantized results $\{\hat{\vb}_k(B)\}_{k=1}^K$ is used as the loss function for the feedback module. This loss function ensures that the VQ codebook $\Qb$ is optimized to minimize the discrepancy between the original and reconstructed channel feature vectors, which can be expressed as
\begin{equation} \label{L_Q}
    \mathcal{L}_Q = \sum_{k=1}^{K} \left\| \vb_k - \hat{\vb}_k(B) \right\|_2^2.
\end{equation}
$\lambda$ is the hyperparameter that balances the tradeoff between sum-rate maximization and quantization accuracy, potentially accelerating convergence.

Through simulations, we observe that the conventional E2E training approach does not effectively train the RVQ codebook, leading to suboptimal performance. To overcome this challenge, we propose a novel progressive training strategy, which divides the training process for an $N$-layer RVQ into $N$ distinct stages. During the first stage, only the first subcodebook $\Qb_{B_1}^1$ and the parameters of the other modules including $\tilde{\Xb}, \mathbf{\Theta}_{F}, \mathbf{\Theta}_W$ are trained. Once this stage is completed, all the trained parameters are saved. In the second stage, the subcodebook of the first layer is frozen, and we continue to train the second tier of the codebook. Meanwhile, parameters such as $\tilde{\Xb}, \mathbf{\Theta}_{F}, \mathbf{\Theta}_W$ are fine-tuned using the saved parameters from the first stage. This iterative process continues for each subsequent stage until all $N$ subcodebooks are fully trained. The detailed steps of this progressive training approach are outlined in Algorithm \ref{alg: progressive training}.

\begin{algorithm}
\caption{The Proposed Progressive Training Strategy}\label{alg: progressive training}
\KwIn{Training dataset $\mathcal{H}$, hyperparameter $\alpha$, $\beta$, $\lambda$}
Initialize $\tilde{\Xb}$, $\mathbf{\Theta}_{F}$, and $\mathbf{\Theta}_W$ randomly\;
Set $\Qb_B = \emptyset$\;
\For{$n = 1,\cdots,N$}{
Initialize $\Qb_B^{n}$ randomly\;
$\Qb_B = \Qb_B \cup \Qb^{n}_{B_n}$\;
\For{$\operatorname{epoch} = 1,2\cdots$}{
Randomly partition $\mathcal{H}$ into $N_{b}$ disjoint batches such that $\mathcal{H} = \bigcup_{t=1}^{N_b} \mathcal{H}_t$\;
\For{$t=1,\cdots,N_b$}{
Compute loss on $\mathcal{H}_t$ based on $\mathcal{L}$ in \eqref{total loss}\;
Update $\tilde{\Xb}$, $\mathbf{\Theta}_{F}$, $\Qb_{B_n}^n$, and $\mathbf{\Theta}_W$ using gradient descent\;
}
}
Freeze $\Qb_{B_n}^{n}$\;
}
\KwOut{Trained $\tilde{\Xb}$, $\mathbf{\Theta}_{F}$, $\Qb_B$, and $\mathbf{\Theta}_W$.}
\end{algorithm}

\section{Scalability Analysis and Practical Implementations of the Proposed DL-Based Transceiver} \label{Sec. V}

This section evaluates the scalability of the proposed DL-based transceiver design and explores its practical implementation to ensure consistent and superior performance across a wide range of system configurations. In practical FDD systems, key parameters such as the feedback bits $B$ and number of users $K$ are inherently time-varying. In subsequent subsections, we will first analyze the scalability of the proposed DL scheme in the context of changing feedback capacities and user counts, and then present tailored solutions to enable its scalability under these varying conditions.

\subsection{Scalability with Feedback Capacity} \label{Sec. V-A}
%In the previously proposed DL-based framework, the RVQ-based feedback modules are jointly trained with the pilot network and the EGAT. However, the number of codewords in the trained RVQ is constrained by the feedback capacity, $B$. Consequently, any change in $B$ necessitates retraining the entire network to adapt to the updated feedback capacity. This retraining process can be both tedious and impractical, particularly in systems where feedback capacity varies across coherence time due to the changing channel conditions. For practical implementations, it is therefore desirable to develop a robust NN capable of operating across a wide range of feedback capacities without requiring retraining.

In this subsection, we analyze the scalability of the proposed DL-based method with respect to feedback capacity. Among the proposed three modules, the feedback module and the EGAT-based precoding design are particularly sensitive to variations in feedback capacity. Fortunately, the proposed RVQ-VAE framework employs a codebook-based VQ method, where the dimension of each codeword remains fixed regardless of the codebook size. This ensures that the reconstructed channel feature maintains a constant dimension, aligning with EGAT’s input requirements and allowing seamless integration. Consequently, the primary challenge in achieving a scalable design for feedback capacity lies in adapting the quantization/dequantization process to handle different feedback capacities without retraining the model. To illustrate this, we consider a trained model with an $N$-tier RVQ, designed for a feedback budget of $B$ feedback bits, where each tier is assigned $\bar{B} = B / N$ bits. In the deployment phase, the actual feedback capacity, denoted as $B_{\rm deploy} \in \mathcal{B}$, may differ with the trained model's budget, leading to two scenarios: $B_{\rm deploy} < B$ and $B_{\rm deploy} > B$. In the following, we introduce two dedicated solutions to address each case in detail.

\subsubsection{Case 1 with $B_{\rm deploy} < B$}
In Case 1, we introduce an efficient codebook shrinkage method that integrates codebook discarding and clustering-based compression to adjust the trained codebook for smaller feedback bit budgets. This approach begins by retaining the first $N_0$ tiers of the trained codebook while discarding the later subcodebooks to match the available feedback budget. The number of retained VQ layers $N_0$ can be determined by expressing $B_{\rm deploy}$ as
\begin{equation} \label{eqn: decompose B_tilde}
    B_{\rm deploy} = N_0 \Bar{B} + B_r, 
\end{equation}
where $N_0 = \lfloor B_{\rm deploy} / \Bar{B} \rfloor$ represents the number of reserved VQ layers, and $B_r < \bar{B}$ denotes the remaining available bits for quantization. To fully utilize the remaining $B_r$ bits, an additional subcodebook of size $2^{B_r}$ is required. This subcodebook is constructed by compressing the $(N_0+1)$-th tier of the trained codebook, denoted as $\Qb_{\bar{B}}^{N_0 + 1}$, which is originally designed to support $\Bar{B}$-bit quantization. Our objective is to compress $\Qb_{\bar{B}}^{N_0 + 1}$ such that it only contains $2^{B_r}$ codewords. We achieve this compression using the differentiable $k$-means (DKM) clustering method \cite{DKM}, which extends conventional $k$-means by incorporating an attention mechanism. This approach effectively reduces the number of codewords in the codebook while preserving its overall structure and quantization performance.

Specifically, we denote the additional subcodebook as $\tilde{\Qb} = [\tilde{\qb}_{1},\tilde{\qb}_{2},\cdots,\tilde{\qb}_{2^{B_r}}]$ without abuse of notation, which is initialized based on $\Qb_{\bar{B}}^{N_0+1}$ via the $k$-means$++$ algorithm \cite{kmeans++}. During the $(m+1)$-th iteration, given the additional codebook $\tilde{\Qb}^{m}$ from the $m$-th iteration, we construct a distance matrix $\Db^{m+1} \in \Rbb^{2^{\Bar{B}} \x 2^{B_r}}$ to store the negative Euclidean distance between the $i$-th codeword in $\Qb_{\bar{B}}^{N_0 + 1}$ and the $j$-th codeword in $\tilde{\Qb}^m$, $\forall i,j$. Thus, the $(i,j)$-th entry of $\Db^{m+1}$ can be defined as
\begin{equation} \label{eqn: D^k_ij}
    \Db^{m+1} (i,j)= - \| \qb_{i}^{N_0 + 1} - \tilde{\qb}_{j}^{m} \|_2, \quad \forall i,j,
\end{equation}
In parallel, we compute the attention matrix $\Ab^{m+1} \in \Rbb^{2^{\Bar{B}} \x 2^{B_r}}$ to store the attention coefficients. The $(i,j)$-th entry of $\Ab^{m+1}$ is then computed using the softmax function as
\begin{equation} \label{eqn: A^k_ij}
    \Ab^{m+1}(i,j) = \dfrac{\operatorname{exp}(\Db^{m+1}(i,j))}{\sum_{l}\operatorname{exp}(\Db^{m+1}(i,l))}, \quad \forall i,j.
\end{equation}
Then, we update $\tilde{\Qb}^{m+1}$ using the constructed attention coefficients. Specifically, the $i$-th codeword of the additional codebook is updated as 
\begin{equation} \label{eqn: update q_tilde_ki}
    \tilde{\qb}_{i}^{m+1} = \dfrac{\sum_{j}\Ab^{m+1}(i,j)\qb_{j}^{N_0 + 1}} {\sum_{j}\Ab^{m+1}(i,j)}, \quad \forall i.
\end{equation}
The algorithm iterates over these steps until a predefined accuracy criterion is satisfied. The detailed procedure is summarized in Algorithm \ref{alg: DKM}.
\begin{algorithm}
\caption{Differential $K$-means Clustering}\label{alg: DKM}
\KwIn{Trained subcodebook of the $(N_0+1)$-th VQ layer $\Qb_B^{N_0+1}$, the target quantization bits $B_r$, and the halting threshold $\epsilon$\;}
Initialize the additional codebook $\tilde{\Qb}^0 \in \Rbb^{D \x 2^{B_r}}$ based on $\Qb_B^{N_0+1}$ using $k$-means$++$ \cite{kmeans++}, $\forall k$\;
\For{$\operatorname{m} = 0,1,2,\cdots$}{
Construct the distance matrix $\Db^{m+1}$ using \eqref{eqn: D^k_ij}\;
Construct the attention coefficient matrix $\Ab^{m+1}$ using \eqref{eqn: A^k_ij}\;
Update the codeword $\tilde{\qb}_{i}^{m+1}$ using \eqref{eqn: update q_tilde_ki}, $\forall i$\;
\If{$ \| \tilde{\Qb}^{m+1} - \tilde{\Qb}^{m} \|_F^2 / \| \tilde{\Qb}^{m} \|_F^2  \leq \epsilon $}{\textbf{break}}
}
\KwOut{Additional codebook $\tilde{\Qb}$.}
\end{algorithm}
By reserving the first $N_0$ tiers of the original codebook and applying the DKM algorithm for codebook compression, we construct a new RVQ codebook as $\Qb_{B_r} = \{\Qb^1_{\bar{B}}, \Qb^2_{\bar{B}}, \cdots, \Qb^{N_0}_{\bar{B}}, \tilde{\Qb}\}$. This modified codebook effectively supports a feedback capacity of $B_{\rm deploy}$ bits without requiring model retraining.

\subsubsection{Case 2 with $B_{\rm delpoy} > B$}
In Case 2, we can introduce additional tiers of subcodebooks to expand the RVQ model. By leveraging the progressive training strategy, only a small subset of parameters needs to be updated, effectively reducing quantization error and enhancing transmission performance with minimal computational cost. Alternatively, model-based codebook expansion methods, such as interpolation or inverse DKM \cite{IDKM}, can also be employed to extend the trained model for larger feedback capacities. These approaches completely avoid model retraining, thereby further reducing model complexity and training overhead.

\subsection{Scalability with Number of Users} \label{Sec. V-B}
In this subsection, we analyze the scalability of the proposed DL-based method with respect to the $K$ across different modules. First, the pilot training network remains scalable since the pilot matrix $\tilde{\Xb}$ is shared among all users. Second, in contrast to \cite{Yu-TWC, GaoZhen_JSAC}, where different users are assigned different feature extractors, quantizers, and dequantizers, our method employs a shared feedback module consisting of a common feature extractor and RVQ codebook for all users. Such a design decouples the feedback mechanism from the user count, facilitating the scalable transceiver design. Third, the EGAT-based precoding network achieves scalability through shared function weights and a message-passing mechanism, making the precoding process independent of $K$. With these designs, the proposed DL-based scheme can accommodate arbitrary user numbers without requiring model retraining, thereby enhancing generalization performance and facilitating practical deployment.
\begin{remark} 
It is worth noting that the feedback module sharing strategy can be naturally extended to accommodate multiple RVQ codebooks. Specifically, multiple codebooks can be designed and dynamically assigned to different users according to certain criteria, such as their geographic locations. For instance, the coverage area of the BS can be partitioned into multiple distinct regions, with each region maintaining its unique codebook. These regional codebooks can then be assigned to users according to their positions within the coverage area. Such an approach not only preserves the scalability but also further improves the feedback performance.
\end{remark}

\section{Numerical Results} \label{Sec. VI}
\subsection{Dataset Generation} \label{Sec. VI-A}
We use Remcom Wireless Insite\footnote{\href{https://www.remcom.com/wireless-insite-propagation-software}{https://www.remcom.com/wireless-insite-propagation-software}} to model a total of $T=128$ physical environments with each measuring $128 \x 128$ square meters with static buildings. The BS is equipped with $M = 128$ antennas in a uniform planar array with dual-polarization, operating at $3.5$GHz. The height of the BS is set as $25$ meters, which is located at the center of each considered area. The antenna spacing is half-wavelength. Each area is uniformly discretized into $16,384$ $1 \x 1$ square meters grids, while the single-antenna users are randomly located at the center of different grids. The height of each user antenna is $1.5$ meters. We generate $S=1,000$ user location realizations based on a uniform distribution. Given the locations of the BS, buildings, and users, we can use the ray-tracing simulator to generate the user channels. Therefore, we have a total of $ST=12,8000$ channel samples. We use $100,000$ samples for training, $18,000$ samples for validation, and $10,000$ samples for testing. 
\begin{figure}
    \centering
    \includegraphics[width=1\linewidth]{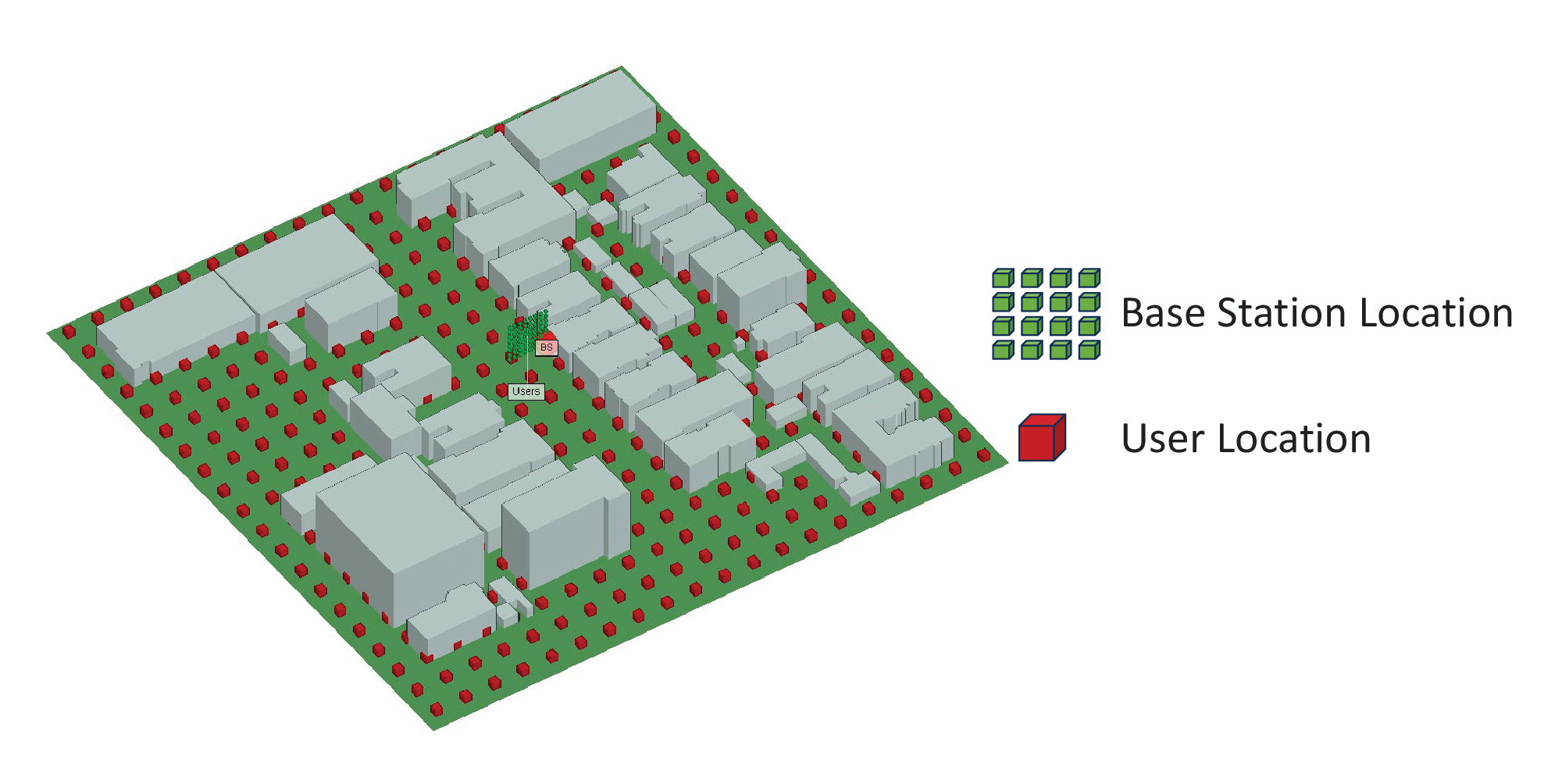}
    \caption{An example of the ray-tracing simulation setup is illustrated. The gray objects represent buildings. The red cubes indicate the positions of single-antenna users, distributed across the simulation area. The green cube array marks the location of the BS.}
    \label{fig: ray-tracing}
\end{figure}

\subsection{Baseline Methods}
To demonstrate the effectiveness of the proposed DL approach, the following baseline methods are considered:
\begin{itemize}
    \item \textbf{ZF: CSIT}: The precoding matrix is obtained using zero-forcing (ZF), assuming the BS has perfect CSI. Specifically, the precoding vector for user $k$, $\wb^{\text{ZF}}_k \in \Cbb^{M \x 1}$, is given by $\wb^{\text{ZF}}_k = \gamma_k \mb_k / \| \mb_k \|_2$, where $\gamma_k = \sqrt{P/K}$ denotes the power allocation factor, and $\mb_k$ is the $k$-th column vector of matrix $\Hb^H (\Hb \Hb^H)^{-1}$.
    \item \textbf{ZF: Imperfect CE \& Perfect Feedback}: The channel vector $\hb_k$ of user $k$ is estimated from the pilot signal with length $L$ using the linear minimum mean squared error (LMMSE) algorithm. Then, user $k$ perfectly feeds back the estimated channel to the BS, which is then utilized to obtain the ZF precoder.
    \item \textbf{ZF: Perfect CE \& Imperfect Feedback}: The channel vectors are assumed to be perfectly known by the users. Then, each user applies the Lloyd algorithm to generate a feedback codebook for quantizing their channels and sending them back to the BS with $B$ bits \cite{Lloyd}. In the simulation, the RVQ codebook with $N$ tiers is used. The construction process begins by applying the Lloyd algorithm to the training dataset to generate the first-tier subcodebook. This subcodebook is then used to quantize the entire training dataset, producing a quantized version of the dataset. The residual between the original and the quantized dataset is computed and used to form a new dataset. The Lloyd algorithm is then applied again to this new dataset to generate the second-tier subcodebook. This process is iteratively repeated until all tiers of the subcodebooks are obtained.
    \item \textbf{DL Method in \cite{Yu-TWC}}: The NN framework proposed in \cite{Yu-TWC} jointly optimizes the pilot generation, user feedback, and precoding strategy, where it adopts a user-specific encoder with a binarization layer to quantize and forward the pilot signal, with a multi-layer FCN at the BS to generate the precoding matrix.
\end{itemize}

\subsection{Setup for NN Parameters and Training Process}
The network is implemented using PyTorch with \textit{Adam} optimizer for training. A cosine annealing learning rate schedule with linear warmup is used to accelerate convergence \cite{Warmup}. The learning rate starts with $10^{-3}$, which gradually decreases to $10^{-5}$. For the hyperparameters, we set $\lambda=1$, $\alpha = \beta = 0.1$, and batch size as $500$. The feature extraction network is a $4$-layer FCN with hidden neurons specified as $\{l_1, l_2, l_3, l_4\} = \{1024, 512, 256, D\}$. Both the initial and normalization layers of the EGAT are implemented with two-layer FCNs, with hidden neurons set as $\{g_1, g_2\} = \{512, 128\}$ and $\{g_1^{\prime}, g_2^{\prime}\} = \{512, 2\}$, respectively. The EGAT employs a three-layer updating mechanism, where each layer’s linear processing function has a dimension of $128$. The \textit{Mish} activation function is applied to all hidden layers except the output layer of the feature extraction network, where \textit{Tanh} is used. Batch normalization is applied after each hidden layer of the feature extraction network to further improve convergence speed \cite{Batch}.\footnote{Our code can be found in \href{https://github.com/LINZHU-PolyU/Scalable-Precoding-MU-MIMO}{https://github.com/LINZHU-PolyU/Scalable-Precoding-MU-MIMO}}

\subsection{Performance Comparison under Different SNR} \label{Sec. Case 1}
\begin{figure}[!t]
\centerline{\includegraphics[width=0.47\textwidth]{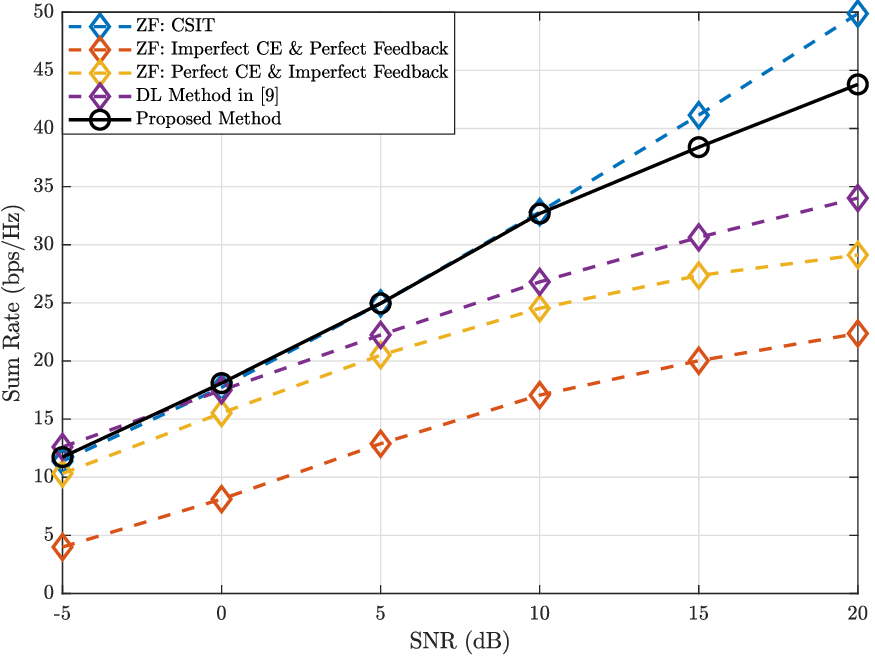}}
\caption{Sum-rate v.s. SNR with $M=128, L=32, K=6, B=30, N = 3$.}
\label{Fig: Rate vs SNR}
\end{figure}

We evaluate the sum-rate performance of the proposed scheme and the baseline methods under different signal-to-noise (SNR) setups, where SNR is defined as $\text{SNR} = 10\log_{10}(P / \sigma^2)$, and the pilot sequence length is $L=32$. The feedback bit budget is set to $B=30$. The RVQ codebook consists of $N=3$ layers with each layer allocated $\Bar{B} = 10$ bits. The proposed model is trained under $\text{SNR} = 20\text{dB}$ and then tested under different SNR regimes. From Fig. \ref{Fig: Rate vs SNR}, it is noted that our proposed scheme achieves similar sum-rate performance as the ZF precoding with full CSIT, which serves as the performance upper bound. Furthermore, by comparing the sum-rate achieved by our scheme and that achieved by the second and the third baseline methods, it is observed that when accurate channel estimation and feedback are not available due to limited resources, the proposed scheme still enables the BS to extract essential features for effective beamforming design. Additionally, the proposed model significantly outperforms the DL-based method in \cite{Yu-TWC}, particularly in high SNR regimes. This shows the strong feature extraction and quantization capabilities of the RVQ-VAE framework and demonstrates the effectiveness of the EGAT-based precoding strategy in improving sum-rate performance.
%We attribute this performance gain to two factors. First, the method in \cite{Yu-TWC} employs a binarization layer to directly map the pilot signals into binary-valued latent features. However, such binarization cannot adequately represent the statistical properties of the actual channel \cite{HoonLee-TCom}. In contrast, the proposed RVQ-based feedback mechanism learns a finite-alphabet channel codebook in a continuous vector space. The learned codebook elements act as continuous-valued channel feature codewords, which can effectively encode the statistical properties of the channel dataset and provide enhanced reconstruction performance. Second, unlike the FCN-based precoding approach in \cite{Yu-TWC}, the proposed EGAT-based precoding design leverages the GNN framework to fully exploit the 2D-PE characteristics inherent in the sum-rate maximization problem. Additionally, the attention mechanism in EGAT dynamically adjusts interference among different users in the network, thereby further improving the sum-rate performance \cite{Yangcy-WCNC}.

\subsection{Scalability Evaluation with Feedback Capacity} \label{Sec. Case 2}
\begin{figure}[!t]
\centerline{\includegraphics[width=0.47\textwidth]{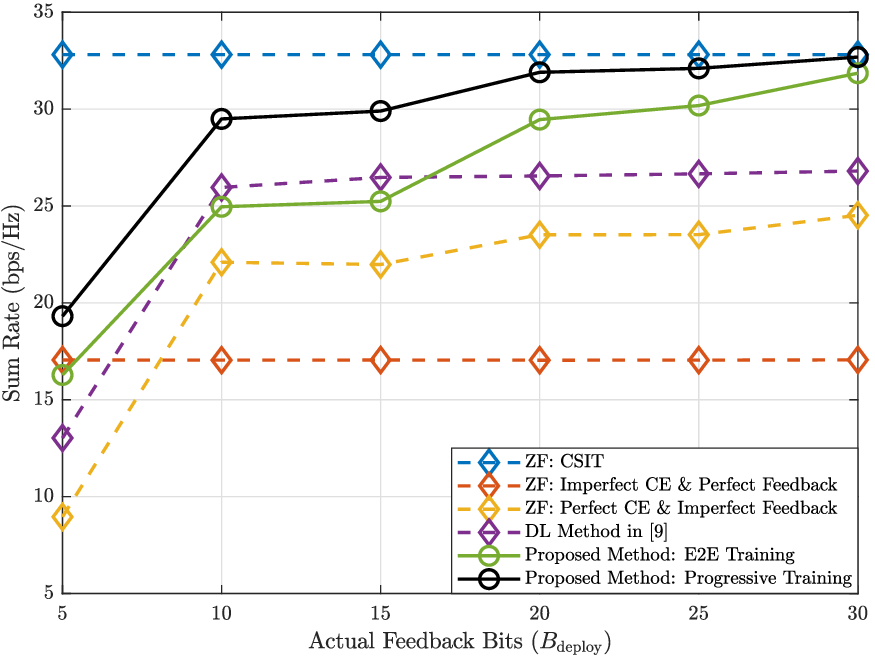}}
\caption{Sum-rate v.s. feedback bit budgets with $M=128, L=32, K=6$.}
\label{Fig: Rate vs B}
\end{figure}

We evaluate the scalability of the proposed framework with respect to varying feedback rates. In this study, the set of all possible feedback bits available to each user is defined as $\mathcal{B} = \{5,10,15,20,25,30\}$. The proposed scheme employs an RVQ codebook with $N=3$ tiers, where each subcodebook is allocated $\bar{B}=10$ bits to support $B=30$-bit quantization. The proposed NN is trained once using the progressive training strategy outlined in Algorithm \ref{alg: progressive training} and is then tested under for each $ B_{\rm deploy} \in \mathcal{B}$, using the proposed codebook shrinkage method. For comparison, we also examine a variant of the proposed model trained using the conventional E2E method, where all RVQ codebook are jointly optimized. In addition, we also consider the DL-based method from \cite{Yu-TWC}, which is trained separately for each $B_{\rm deploy} \in \mathcal{B}$. In the simulation, the SNR is set to $10\text{dB}$. The pilot length and the number of users are specified as $L=32$ and $K=6$, respectively. As illustrated in Fig. \ref{Fig: Rate vs B}, the proposed scheme with progressive training achieves performance close to ZF with perfect CSIT when $B_{\rm deploy} \geq 20$. It also significantly outperforms conventional methods and the DL-based model from \cite{Yu-TWC} across all tested feedback capacities. Furthermore, the progressive training approach consistently surpasses the E2E training strategy. These facts highlight the effectiveness of the proposed scheme and training policy in enhancing scalability concerning feedback capacity while improving overall sum-rate performance.

\subsection{Scalability Evaluation with User Number} \label{Sec. Case 3}
\begin{figure}[!t]
\centerline{\includegraphics[width=0.47\textwidth]{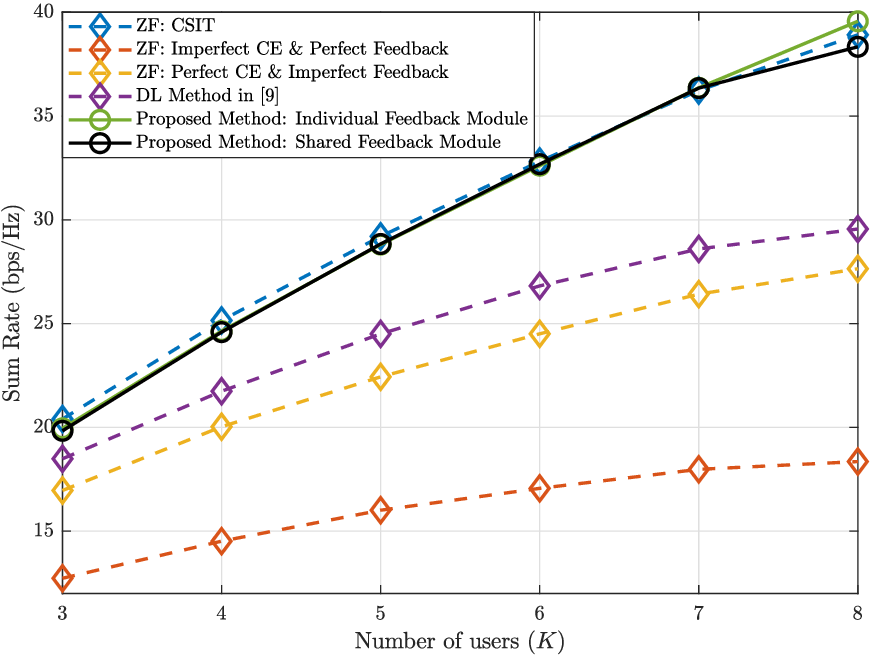}}
\caption{Sum-rate v.s. number of users with $M=128, L=32, B=30, N=3$.}
\label{Fig: Rate vs K}
\end{figure}

To assess the scalability of the proposed scheme with varying numbers of users, we define the set of all the possible scheduled users as $\mathcal{K} = \{3,4,5,6,7,8\}$. For the proposed DL framework with the feedback module sharing strategy, a single NN is trained using a dataset in which each training sample contains randomly selected $K \in \mathcal{K}$ users. For comparison, we consider a variant of our scheme where each user is assigned a dedicated feedback module. Unlike our approach, this variant and the DL-based method from \cite{Yu-TWC} require separate training and parameter storage for each $K \in \mathcal{K}$. In the simulation, the SNR is set to $10\text{dB}$, with a pilot sequence length of $L=32$ and a feedback bit budget of $B=30$. The number of RVQ stages is set as $N = 3$. As shown in Fig. \ref{Fig: Rate vs K}, the sum-rate performance of the proposed scheme with a shared feedback module still significantly outperforms the conventional schemes. Notably, as $K$ increases, the performance loss of the scheme proposed in \cite{Yu-TWC} to our proposed scheme becomes more significant. Furthermore, the performance gap between the shared feedback module and the individual feedback module variant remains minimal, indicating that the proposed framework effectively scales with the number of users while maintaining low model complexity and minimal performance degradation.

\subsection{Impact of RVQ Stages on Quantization Complexity and Sum-Rate Performance} \label{Sec. Case 4}
\begin{figure}[!t]
\centerline{\includegraphics[width=0.47\textwidth]{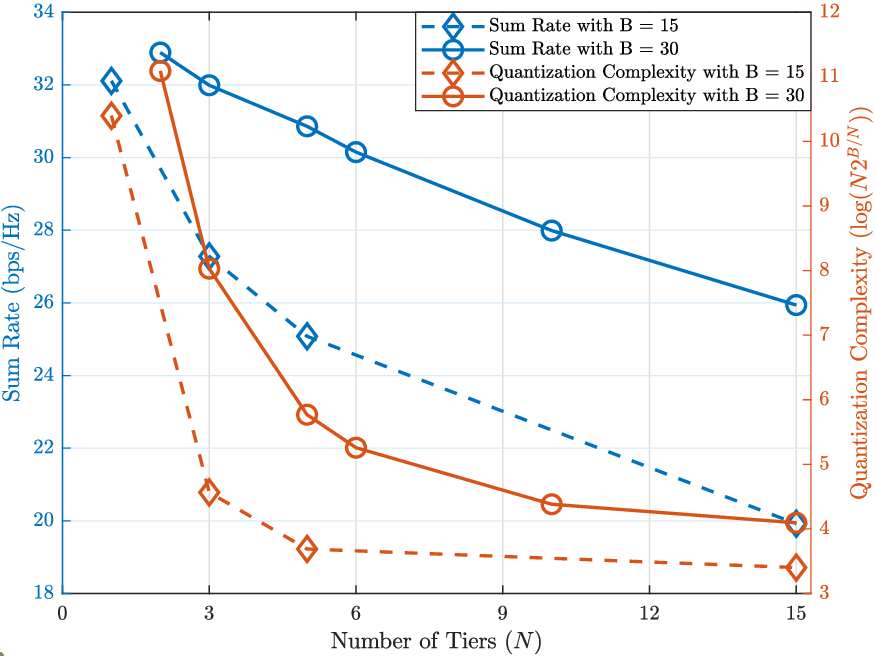}}
\caption{Comparison of number of VQ stages, quantization complexity, and sum-rate performance with $M=128, L=32, K=6$.}
\label{Fig: Rate vs Bit Allocation}
\end{figure}

In this subsection, we analyze the impact of the number of RVQ stages on quantization complexity and sum-rate performance. The proposed framework adopts an equal bit allocation strategy across all VQ stages, where the total number of codewords is given by $N2^{B/N}$. Therefore, the number of VQ stages $N$ plays a crucial role in balancing computational complexity and reconstruction accuracy. To quantify this tradeoff, we characterize the quantization complexity using $\log(N2^{B/N})$. The simulation is conducted with an $\rm SNR$ of $10\rm dB$, a pilot length of $L=32$, and $K = 6$ users, under two different feedback bit budgets: $B=15$ and $B = 30$. As illustrated in Fig. \ref{Fig: Rate vs Bit Allocation}, configurations with fewer VQ stages (each having a larger subcodebook) achieve higher sum-rate performance but significantly increase computational complexity. However, by carefully selecting $N$, it is possible to substantially reduce complexity while maintaining strong performance. For instance, with $B=15$ and $N = 3$, the configuration retains $84.96\%$ of the sum-rate performance of the case with $N=1$, while using only $0.29\%$ of the codewords. Similarly, when $B = 30$ and $N = 5$, the configuration achieves $93.83\%$ of the sum-rate performance of the case when $N = 2$, while using just $0.49\%$ of the codewords. These results highlight the effectiveness of RVQ in balancing quantization complexity and sum-rate performance.
%\begin{table}[h]
%    \centering
%    \caption{Tradeoff between the quantization complexity and sum-rate performance when $B = 15$ and $B = 30$.}
%    \label{tab: Bit allocation}
%    \renewcommand{\arraystretch}{1.2}
%    \begin{tabular}{c|c|c|c}
%        \hline\hline
%        Setup & Number of layers & Codebook size & Sum-rate \\
%        \hline\hline
%        \multirow{4}{*}{$B=15$} & $1$ & $32768$ & $32.11$ \\
%        & $3$ & $96$ & $27.28$ \\
%        & $5$ & $40$ & $25.08$ \\
%        & $15$ & $30$ & $19.92$ \\
%        \hline
%        \multirow{6}{*}{$B=30$} & $2$ & $65536$ & $32.89$ \\
%        & $3$ & $3072$ & $31.99$ \\
%        & $5$ & $320$ & $30.86$ \\
%        & $6$ & $192$ & $30.15$ \\
%        & $10$ & $80$ & $27.99$ \\
%        & $15$ & $60$ & $25.94$ \\
%        \hline
%    \end{tabular}
%\end{table}

\subsection{Generalization Performance to Unseen Environments} \label{Sec. Case 5}
In this subsection, For training, we continue to use the dataset generated by ray-tracing as obtained from Section \ref{Sec. VI-A}. However, for testing, we include $5$ additional unseen environments, which are also modeled using ray-tracing. For each unseen environment, $1,000$ channel samples are generated for testing. The simulation is conducted with an $\rm SNR$ of $10\rm dB$, a pilot sequence length of $L=32$, $K=6$ users, a feedback bit budget of $B=30$, and $N=3$ RVQ stages. As shown in Table \ref{tab: Unseen environments}, the proposed DL-based method continues to significantly outperform conventional schemes, even on these previously unseen environments. These results highlight that the proposed architecture maintains strong performance without requiring retraining for new environments, further confirming its practical feasibility for real-world applications.
\begin{table}[h]
    \centering
    \caption{Sum-rate evaluation of the proposed method for unseen environments with $M=128, L=32, K=6, B=30, N=3$}
    \label{tab: Unseen environments}
    \renewcommand{\arraystretch}{1.2}
    \begin{tabular}{c|c|c|c}
        \hline\hline
        \multirow{2}{*}{Areas} & Proposed & ZF: Imperfect CE & ZF: Perfect CE \\
        & Method & $\&$ Perfect Feedback & $\&$ Imperfect Feedback \\
        \hline\hline
        Area 1 & $23.01$ & $15.40$ & $17.48$ \\
        Area 2 & $17.14$ & $12.56$ & $13.87$ \\
        Area 3 & $14.34$ & $9.86$ & $11.28$ \\
        Area 4 & $10.55$ & $8.08$ & $7.88$ \\
        Area 5 & $16.63$ & $13.71$ & $13.59$ \\
        \hline
    \end{tabular}
\end{table}

\section{Conclusion} \label{Sec. VII} 

In this paper, we proposed a novel DL-based framework for scalable transceiver design in practical FDD massive MIMO systems. The framework comprised a dedicated NN for optimizing pilot signals, a novel RVQ-VAE-based feedback module for efficient channel feedback, and an EGAT for robust multi-user precoding. The hierarchical structure of the RVQ allowed for flexible adjustment of the codebook size, enabling the system to accommodate varying feedback capacities. Additionally, the feedback module sharing strategy and the inherent scalability of the EGAT-based multi-user precoding enabled the framework to effectively scale with the number of users. To further enhance performance, we introduced a progressive training strategy that updated the RVQ codebooks sequentially. Numerical results based on a real-world channel dataset demonstrated the effectiveness of our framework and its generalization capability to unseen environments.

\ifCLASSOPTIONcaptionsoff
  \newpage
\fi

%\begin{thebibliography}{1}
\bibliographystyle{IEEEtran}
\bibliography{IEEEabrv.bib, reference.bib}

\begin{thebibliography}{10}
\providecommand{\url}[1]{#1}
\csname url@samestyle\endcsname
\providecommand{\newblock}{\relax}
\providecommand{\bibinfo}[2]{#2}
\providecommand{\BIBentrySTDinterwordspacing}{\spaceskip=0pt\relax}
\providecommand{\BIBentryALTinterwordstretchfactor}{4}
\providecommand{\BIBentryALTinterwordspacing}{\spaceskip=\fontdimen2\font plus
\BIBentryALTinterwordstretchfactor\fontdimen3\font minus
  \fontdimen4\font\relax}
\providecommand{\BIBforeignlanguage}[2]{{%
\expandafter\ifx\csname l@#1\endcsname\relax
\typeout{** WARNING: IEEEtran.bst: No hyphenation pattern has been}%
\typeout{** loaded for the language `#1'. Using the pattern for}%
\typeout{** the default language instead.}%
\else
\language=\csname l@#1\endcsname
\fi
#2}}
\providecommand{\BIBdecl}{\relax}
\BIBdecl

\bibitem{Larsson_MCOM}
E.~G. Larsson, O.~Edfors, F.~Tufvesson, and T.~L. Marzetta, ``Massive {MIMO}
  for next generation wireless systems,'' \emph{{IEEE} Commun. Mag.}, vol.~52,
  no.~2, pp. 186--195, Feb. 2014.

\bibitem{Lulu_JSTSP}
L.~Lu, G.~Y. Li, A.~L. Swindlehurst, A.~Ashikhmin, and R.~Zhang, ``An overview
  of massive {MIMO}: Benefits and challenges,'' \emph{{IEEE} J. Sel. Topics
  Signal Process.}, vol.~8, no.~5, pp. 742--758, Oct. 2014.

\bibitem{Jindal-TIT}
N.~Jindal, ``{MIMO} broadcast channels with finite-rate feedback,''
  \emph{{IEEE} Trans. Inf. Theory}, vol.~52, no.~11, pp. 5045--5060, Nov. 2006.

\bibitem{Jindal-JSAC}
N.~Ravindran and N.~Jindal, ``Limited feedback-based block diagonalization for
  the {MIMO} broadcast channel,'' \emph{{IEEE} J. Sel. Areas Commun.}, vol.~26,
  no.~8, pp. 1473--1482, Oct. 2008.

\bibitem{XiaPengfei_TSP}
P.~Xia and G.~Giannakis, ``Design and analysis of transmit-beamforming based on
  limited-rate feedback,'' \emph{{IEEE} Trans. Signal Process.}, vol.~54,
  no.~5, pp. 1853--1863, May 2006.

\bibitem{Yu-TWC}
F.~Sohrabi, K.~M. Attiah, and W.~Yu, ``Deep learning for distributed channel
  feedback and multiuser precoding in {FDD} massive {MIMO},'' \emph{{IEEE}
  Trans. Wireless Commun.}, vol.~20, no.~7, pp. 4044--4057, Jul. 2021.

\bibitem{GaoZhen_JSAC}
Z.~Gao \emph{et~al.}, ``Data-driven deep learning based hybrid beamforming for
  aerial massive {MIMO-OFDM} systems with implicit {CSI},'' \emph{{IEEE} J.
  Sel. Areas Commun.}, vol.~40, no.~10, pp. 2894--2913, Oct. 2022.

\bibitem{HoonLee-TCom}
J.~Jang, H.~Lee, I.-M. Kim, and I.~Lee, ``Deep learning for multi-user {MIMO}
  systems: Joint design of pilot, limited feedback, and precoding,''
  \emph{{IEEE} Trans. Commun.}, vol.~70, no.~11, pp. 7279--7293, Nov. 2022.

\bibitem{Caire-ICC}
Y.~Song, T.~Yang, M.~B. Khalilsarai, and G.~Caire, ``Deep-learning aided
  channel training and precoding in {FDD} massive {MIMO} with channel
  statistics knowledge,'' in \emph{Proc. {IEEE} Int. Conf. Commun. ({ICC})},
  2023, pp. 2791--2797.

\bibitem{Carpi-ICC}
F.~Carpi \emph{et~al.}, ``Precoding-oriented massive {MIMO} {CSI} feedback
  design,'' in \emph{Proc. IEEE Int. Conf. Commun. ({ICC})}, 2023, pp.
  4973--4978.

\bibitem{HuangHongji}
H.~Huang, J.~Yang, H.~Huang, Y.~Song, and G.~Gui, ``Deep learning for
  super-resolution channel estimation and {DOA} estimation based massive {MIMO}
  system,'' \emph{{IEEE} Trans. Veh. Technol.}, vol.~67, no.~9, pp. 8549--8560,
  Sep. 2018.

\bibitem{DongPeihao}
P.~Dong, H.~Zhang, G.~Y. Li, I.~S. Gaspar, and N.~NaderiAlizadeh, ``Deep
  {CNN}-based channel estimation for mmwave massive {MIMO} systems,''
  \emph{{IEEE} J. Sel. Topics Signal Process.}, vol.~13, no.~5, pp. 989--1000,
  Sep. 2019.

\bibitem{CSINet}
C.-K. Wen, W.-T. Shih, and S.~Jin, ``Deep learning for massive {MIMO} {CSI}
  feedback,'' \emph{{IEEE} Commun. Lett.}, vol.~7, no.~5, pp. 748--751, Oct.
  2018.

\bibitem{XiaWenchao}
W.~Xia, G.~Zheng, Y.~Zhu, J.~Zhang, J.~Wang, and A.~P. Petropulu, ``A deep
  learning framework for optimization of {MISO} downlink beamforming,''
  \emph{{IEEE} Trans. Commun.}, vol.~68, no.~3, pp. 1866--1880, Mar. 2020.

\bibitem{Bruno_GLOBECOM}
B.~Clerckx, G.~Kim, J.~Choi, and S.~Kim, ``Allocation of feedback bits among
  users in broadcast {MIMO} channels,'' in \emph{Proc. of {IEEE} Global Commun.
  Conf. ({GLOBECOM})}, Dec. 2008, pp. 1--5.

\bibitem{LeeJungHoon_TWC}
J.~H. Lee and W.~Choi, ``Optimal feedback rate sharing strategy in zero-forcing
  {MIMO} broadcast channels,'' \emph{{IEEE} Trans. Wireless Commun.}, vol.~12,
  no.~6, pp. 3000--3011, Jun. 2013.

\bibitem{KB_TWC}
B.~Khoshnevis and W.~Yu, ``Bit allocation laws for multiantenna channel
  feedback quantization: {Multiuser} case,'' \emph{{IEEE} Trans. Signal
  Process.}, vol.~60, no.~1, pp. 367--382, Jan. 2012.

\bibitem{GuoJiajia_TWC}
J.~Guo, C.-K. Wen, S.~Jin, and G.~Y. Li, ``Convolutional neural network-based
  multiple-rate compressive sensing for massive {MIMO} {CSI} feedback: Design,
  simulation, and analysis,'' \emph{{IEEE} Trans. Wireless Commun.}, vol.~19,
  no.~4, pp. 2827--2840, Apr. 2020.

\bibitem{LinYuchien_TWC}
Y.-C. Lin, T.-S. Lee, and Z.~Ding, ``A scalable deep learning framework for
  dynamic {CSI} feedback with variable antenna port numbers,'' \emph{{IEEE}
  Trans. Wireless Commun.}, vol.~23, no.~4, pp. 3102--3116, Apr. 2024.

\bibitem{Nerini_TWC}
M.~Nerini, V.~Rizzello, M.~Joham, W.~Utschick, and B.~Clerckx, ``Machine
  learning-based {CSI} feedback with variable length in {FDD} massive {MIMO},''
  \emph{{IEEE} Trans. Wireless Commun.}, vol.~22, no.~5, pp. 2886--2900, May
  2023.

\bibitem{Shenyifei_TWC}
Y.~Shen, J.~Zhang, S.~H. Song, and K.~B. Letaief, ``Graph neural networks for
  wireless communications: From theory to practice,'' \emph{{IEEE} Trans.
  Wireless Commun.}, vol.~22, no.~5, pp. 3554--3569, May 2023.

\bibitem{YangCY_TWC}
S.~Liu, J.~Guo, and C.~Yang, ``Multidimensional graph neural networks for
  wireless communications,'' \emph{{IEEE} Trans. Wireless Commun.}, vol.~23,
  no.~4, pp. 3057--3073, Apr. 2024.

\bibitem{Yu-JSAC}
T.~Jiang, H.~V. Cheng, and W.~Yu, ``Learning to reflect and to beamform for
  intelligent reflecting surface with implicit channel estimation,''
  \emph{{IEEE} J. Sel. Areas Commun.}, vol.~39, no.~7, pp. 1931--1945, Jul.
  2021.

\bibitem{Rizzello_ICC}
V.~Rizzello, D.~B. Amor, M.~Joham, and W.~Utschick, ``Scalable multi-user
  precoding and pilot optimization with graph neural networks,'' in \emph{Proc.
  {IEEE} Int. Conf. Commun. ({ICC})}, 2024, pp. 2956--2961.

\bibitem{Taomx-Lett}
J.~Yang \emph{et~al.}, ``Deep learning for joint design of pilot, channel
  feedback, and hybrid beamforming in {FDD} massive {MIMO-OFDM} systems,''
  \emph{{IEEE} Commun. Lett.}, vol.~28, no.~2, pp. 313--317, Feb. 2024.

\bibitem{VQVAE}
A.~van~den Oord, O.~Vinyals, and K.~Kavukcuoglu, ``Neural discrete
  representation learning,'' in \emph{Proc. Adv. Neural Inf. Process. Syst.},
  2017, pp. 6306--6315.

\bibitem{NSVQ}
M.~H. Vali and T.~B{\"a}ckstr{\"o}m, ``{NSVQ}: Noise substitution in vector
  quantization for machine learning,'' \emph{{IEEE} Access}, pp.
  13\,598--13\,610, 2022.

\bibitem{Yangcy-WCNC}
S.~Liu and C.~Yang, ``Learning user scheduling and hybrid precoding with
  sequential graph neural network,'' in \emph{Proc. {IEEE} Wireless Commun.
  Netw. Conf. ({WCNC})}, 2024, pp. 1--6.

\bibitem{DKM}
M.~Cho, K.~A. Vahid, S.~Adya, and M.~Rastegari, ``{DKM}: Differentiable k-means
  clustering layer for neural network compression,'' 2022,
  \emph{arXiv:2108.12659}. [Online]. Available:
  \href{https://arxiv.org/abs/2108.12659}{https://arxiv.org/abs/2108.12659}.

\bibitem{kmeans++}
D.~Arthur and S.~Vassilvitskii, ``{k-means++}: The advantages of careful
  seeding,'' in \emph{Proc. 18th Ann. ACM-SIAM Symp. Discrete Algorithms
  {(SODA)}}, 2007, pp. 1027--1035.

\bibitem{IDKM}
J.~Seo and J.~Kang, ``Rate-adaptive quantization: {A} multi-rate codebook
  adaptation for vector quantization-based generative models,'' 2025,
  \emph{arXiv:2405.14222}. [Online]. Available:
  \href{https://arxiv.org/abs/2405.14222}{https://arxiv.org/abs/2405.14222}.

\bibitem{Lloyd}
A.~Gersho and R.~M. Gray, \emph{Vector Quantization and Signal
  Compression}.\hskip 1em plus 0.5em minus 0.4em\relax Norwell, MA: Kluwer,
  1992.

\bibitem{Warmup}
I.~Loshchilov and F.~Hutter, ``{SGDR}: Stochastic gradient descent with warm
  restarts,'' 2017, \emph{arXiv:1608.03983}. [Online]. Available:
  \href{https://arxiv.org/abs/1608.03983}{https://arxiv.org/abs/1608.03983}.

\bibitem{Batch}
S.~Ioffe and C.~Szegedy, ``Batch normalization: Accelerating deep network
  training by reducing internal covariate shift,'' 2015,
  \emph{arXiv:1502.03167}. [{Online}]. Available:
  \href{https://arxiv.org/abs/1502.03167}{https://arxiv.org/abs/1502.03167}.

\end{thebibliography}

%\end{thebibliography}

\end{document}